\documentclass[11pt,letterpaper,english]{amsart}
\usepackage[utf8]{inputenc}
\usepackage[letterpaper, top=1in, bottom=1in, left=1in, right=1in]{geometry}

\usepackage[foot]{amsaddr}

\newcommand{\comment}[1]{}
\usepackage{mathtools}
\usepackage[T1]{fontenc}
\usepackage{amssymb}
\usepackage[english]{babel}
\usepackage{float}
\usepackage{enumerate}
\usepackage{color}
\usepackage{comment}
\usepackage{amsmath}
\usepackage{tikz}
\usepackage{url}
\usepackage[ruled,vlined]{algorithm2e}

\usepackage{amsthm}
\usepackage{multirow}
\usepackage{booktabs}
\usepackage{thmtools}
\usepackage{dsfont}

\usepackage{afterpage}
\usepackage{capt-of}

\usepackage{hyperref}
\usepackage[capitalise, noabbrev]{cleveref}

\crefname{@theorem}{Theorem}{Theorems}

\usepackage{thmtools,thm-restate}

\theoremstyle{plain}
\newtheorem{thm}{Theorem}
\newtheorem{lem}[thm]{Lemma}

\theoremstyle{definition}

\theoremstyle{remark}

\newtheoremstyle{break}
  {}
  {}
  {\itshape}
  {}
  {\bfseries}
  {.}
  {\newline}
  {}

\theoremstyle{break}

\crefname{lem}{Lemma}{lemmas}

\usepackage{todonotes}

\newcommand{\ALG}{\textsc{ALG}}
\newcommand{\OPT}{\textsc{OPT}}
\newcommand{\PP}{\mathbb{P}}
\newcommand{\E}{\mathbb{E}}
\newcommand{\Y}{\mathcal{Y}}

\newcommand{\CLP}{\mathrm{CLP}}
\newcommand{\SDCLP}{\mathrm{SDCLP}}
\newcommand{\UBP}{\mathrm{UBP}}
\newcommand{\LBP}{\mathrm{LBP}}
\newcommand{\LP}{\mathrm{LP}}
\newcommand{\SDLP}{\mathrm{SDLP}}
\newcommand{\SDRP}{\mathrm{SDRP}}

\newcommand{\RP}{\mathrm{RP}}

\newcommand{\A}{\mathcal{A}}

\newcommand{\R}{\mathbb{R}}
\newcommand{\N}{\mathbb{N}}

\DeclarePairedDelimiter\floor{\lfloor}{\rfloor}

\title[Sample-driven optimal stopping]{Sample-driven optimal stopping: From the secretary problem to the i.i.d.~prophet inequality}

 \author[J. Correa]{José Correa$^1$}
 \author[A. Cristi]{Andr\'es Cristi$^1$}
 \address{$^1$Universidad de Chile}
 \author[B. Epstein]{Boris Epstein$^2$}
 \address{$^2$Columbia University}
  \author[J.A. Soto]{Jos\'e A. Soto$^1$}

 \date{July 2021}

\begin{document}

\maketitle
\thispagestyle{empty}
\begin{abstract}
We take a unifying approach to single selection optimal stopping problems with random arrival order and independent sampling of items. In the problem we consider, a decision maker (DM) initially gets to sample each of $N$ items independently with probability $p$, and can observe the relative rankings of these sampled items. Then, the DM faces the remaining items in an online fashion, observing the relative rankings of all revealed items. While scanning the sequence the DM makes irrevocable stop/continue decisions and her reward for stopping the sequence facing the item with rank $i$ is $Y_i$. The goal of the DM is to maximize her reward. We start by studying the case in which the values $Y_i$ are known to the DM, and then move to the case in which these values are adversarial.

For the former case, we write the natural linear program that captures the performance of an algorithm, and take its continuous limit by infinitely extending the $Y$ sequence. Then we prove a structural result about this continuous limit, based on optimal transport, which allows us to reduce the problem of finding the optimal algorithm to a relatively simple real optimization problem. As a consequence we extend previous results by establishing that the optimal algorithm for the problem is given by a sequence of thresholds $t_1\le t_2\le\cdots$ such that the DM should stop if seeing an item with current ranking $i$ after time $t_i$. Additionally we are able to recover several classic results in the area such as those for secretary problem, the 1-choice 2-best  secretary problem, and the minimum ranking problem, thus giving a unifying framework for single selection optimal stopping. 

Then, we turn to study the case in which the $Y$ values are chosen by an adversary. This situation adds one layer to the linear programming approach of the previous case. By means of von Neumann's minmax Theorem we establish  that it is equivalent, from a worst case analysis perspective, if the adversary chooses the values before or after the algorithm is fixed (so we may assume that the values are arbitrary and unknown to the DM). Interestingly, this still leads to a similar linear program than when the $Y$ values are known but with an additional stochastic dominance constraint. By using the same machinery we are able to pin down the optimal algorithm for this problem, obtaining the optimal  competitive ratios for all values of $p$.  Notably, we prove that as $p$ approaches 1, our guarantee approaches 0.745, matching that of the i.i.d.~prophet inequality. This implies that there is no loss by considering this more general combinatorial version without full distributional knowledge. Furthermore, we prove that this convergence is very fast. Also interesting is the case $p=1/2$, as it corresponds to the situation in which the sets of observed an unobserved values are of roughly equal size. Here our bound evaluates to $0.671$, which improves upon the state of the art (for large values of $N$). To wrap-up the paper, we explore consequences of our results beyond single selection problems.
\end{abstract}

\newpage
\clearpage
\pagenumbering{arabic} 

\section{Introduction}

Two fundamental models in online decision making are that of competitive analysis and that of optimal stopping. In the former the input is produced by an \emph{adversary} whose goal is to make the algorithm perform poorly with respect to a certain benchmark. In the latter the algorithm has full distributional knowledge of the input making it much easier for the algorithm to achieve good approximation ratios. The area of optimal stopping has been very active in the last decade since many real-world situations, including several e-commerce platforms, often do not behave adversarially, and the distributional model of optimal stopping seems appropriate. Furthermore, the activity in the area has been boosted by the close connection between posted price mechanisms, attractive for their usability and simplicity, and prophet inequalities, a classic topic in optimal stopping theory \cite{HKS07,CHMS10}. 

Recently, data-driven versions of optimal stopping problems have been successfully studied. These constitute a bridge between the worst case model and the distributional model. A standard model, first described in Azar et al.'s \cite{AKW14} pioneering work, consists in replacing the full distributional knowledge with having access to one or more samples from each distribution. The model is very attractive both from a practical and theoretical perspective. On the one hand, full distributional knowledge is a strong assumption, while access to historical data is usually straightforward. And this historical data can be thought of as being samples from certain underlying distributions. On the other hand, the model gains back the combinatorial flavor of competitive analysis and thus becomes much more prone to be analyzed using standard algorithmic tools. A notable example of this is the recent result of Rubinstein et al.~\cite{RWW20} for the classic prophet inequality \cite{KS77,KS78}. They show that access to a single sample from each distribution is enough to guarantee the best possible factor in the full information case (with adversarial order), namely $1/2$. Inspired by Azar et al.'s model, Correa et al.~\cite{CDFS19} considered the situation in which $n$ i.i.d.~samples, drawn from an unknown distribution, are sequentially presented to a decision maker (DM) who has to select a single value making irrevocable stop/continue decisions. They establish that when the DM has no additional information the best she can do is to basically apply the classic algorithm for the secretary problem and thus obtain, in expectation, a fraction $1/e$ of the expected maximum value. On the other hand if she has access to $O(n^2/\varepsilon)$ samples then she can essentially learn the distribution and guarantee a factor of $\alpha^*-O(\varepsilon)$; where $\alpha^*\approx 0.745$ is the optimal factor for the i.i.d.~prophet inequality with full distributional knowledge \cite{HK82,K86,CFHOV17,LPPSS20}. This latter result was improved by Rubinstein et al.~\cite{RWW20}, who showed that $O(n/\varepsilon^6)$ samples are enough to guarantee a factor of $\alpha^*-O(\varepsilon)$. The sampling model from i.i.d. random variables \cite{CDFS19} shares some aspects with the classic secretary problem, in which arbitrary non-negative numbers are presented to the DM in uniform random order \cite{L61,D63,F89,GM66}. Along these lines, a particularly clean model \cite{CS81,KNR20} is the \emph{dependent} sampling model in which the instance, consisting of $N$ items, is designed by an adversary. Then, the DM gets to sample $h=pN$ of these and scans the remaining items in random order. This model is very robust since it generalizes the sampling model from i.i.d.~random variables while making no distributional assumptions. A closely related sampling model, and essentially equivalent for large values of $N$, is that with \emph{independent} sampling \cite{CCFOT21}. Here, rather than sampling exactly $h=pN$ items, the DM samples each item independently with probability $p$.

In this paper, we study a generic version of the classic single selection optimal stopping problem with sampling, which we call $p$-sample-driven optimal stopping problem ($p$-DOS). In this problem a collection of $N$ items is shuffled in uniform random order. The decision maker gets to observe each item independently with probability $p\in [0,1)$ and these items conform the \emph{information set} or \emph{history set}. The remaining items, conforming the \emph{online set} are revealed sequentially to the DM. At any point, the DM observes the relative rankings of the items that have been revealed, and upon seeing an item, she must decide whether to take it and stop the sequence, or to drop it and continue with the next item. If the DM stops with the $i$-th ranked item she gets a reward of $Y_i$ and her goal is to maximize the expected value with which she stops. While we do assume that the values are monotone, i.e., $Y_1\ge \cdots \ge Y_N$, we do not assume that they are non-negative. The natural benchmark to measure the performance of an algorithm here is the expected (over the permutations) maximum value in the online set.

We study both, the cases when the values $Y$ are fixed, ($p$-DOS with known values), and that when they are adversarial ($p$-DOS with adversarial values). The former, and already when $p$=0, models most well known single selection optimal stopping problems. Indeed the classic secretary problem \cite{L61,D63} appears when $Y_1=1$ and the remaining values are $0$, the $1$-choice $K$-best secretary problem \cite{G66,CCJ15} is recovered by $Y_1=\cdots=Y_K=1$ and filling zeros in the remaining values, while the problem of selecting an item of minimum ranking \cite{CMRS64} is obtained by setting $Y_i=-i$.
Still in the case $p=0$, the problem with non-negative values was studied by Mucci \cite{M73a}. By analyzing the underlying recursion he obtains a limiting ODE and established that the optimal algorithm takes the form of a sequence of thresholds such that starting at time $t_i$ the DM should stop with an item currently ranked $i$ or better. Bearden et al. \cite{BRM06} also consider this problem from an experimental viewpoint, while Mucci \cite{M73b} studies the case in which all $Y$'s are negative.

\subsection*{Our results and overview of the paper}
In this paper we derive the optimal algorithms for problem $p$-DOS with known values and for $p$-DOS with adversarial values, for all $p\in [0,1)$.

After some preliminary definitions in \cref{sec:problem}, 
we start with the case in which $Y$ is known to the DM. Here we take the, by now classic, linear programming approach of Buchbinder, Jain, and Singh \cite{BJS14} though slightly extending it to make it able to deal with arbitrary $Y$ values and sampling probability $p$, and adding a term that forces the algorithm to stop.\footnote{This is needed since some $Y_i$'s may be negative.} We note that this LP exactly encodes the best possible algorithm for the problem and that its objective function value decreases with the number of items $N$ (\cref{sec:LPK}). This allows us to deduce that the hardest instances appear as $N\to \infty$. Thus, in \cref{sec:LimitK}, we derive the limit LP which shares some aspects with that of Chan, Chen, and Jiang \cite{CCJ15}. In uncovering the structure of this limit LP, we provide our first main technical contribution in \cref{sec:structure}. By understanding monotonicity properties of the LP coefficients and by using mass moving arguments from the theory of optimal transport, we can deduce exactly which inequalities, and in what ranges, are tight in an optimal solution. This permits to bring down the problem of finding the optimal algorithm to that of solving certain, very simple, ODEs\footnote{Which are very different to that of Mucci \cite{M73a}.}. We find the explicit solution of these ODEs and thus bring the problem real optimization problem in which the variables are some $t_i$'s determining the ranges where the solutions of the different ODEs should be used. These $t_i$'s also have a natural algorithmic interpretation. They represent the \emph{times} at which the DM should start accepting an item of rank $i$ or higher (among the items seen son far). With this we can conclude that Mucci's structural result holds even if some (or all) $Y$ values are negative and for arbitrary $p$. 

Pushing things a bit further we prove, in \cref{sec:thresholds}, that this optimization problem over $t_i$'s is concave in each variable and relatively easy to solve, at least approximately. In particular we exemplify that its first order conditions quickly allow us to recover the known results for the secretary problem \cite{L61}, the 1-choice 2-best secretary problem \cite{CCJ15}, and the minimum rank problem \cite{CMRS64}.

Then we move to our main contribution; the study of $p$-DOS with adversarial values, which we require to be non-negative. This essentially consists in adding a minimization over $Y$ to the linear program for $p$-DOS with known values. However, to make the problem well posed we first need to normalize the objective function. This is done dividing the objective by the expected value of the optimal choice in the online set, namely $\sum_{i=1}^\infty Y_i(1-p)p^{i-1}.$\footnote{Note that for $p=0$ this is just $Y_1$.} Equivalently, we may add a constraint to the LP imposing that this value is 1. In either way the resulting objective function represents the performance guarantee of an optimal online algorithm. With this formulation, von Neumann's minmax Theorem allows us to rewrite the minmax problem as a new linear program in which the constraints take a stochastic dominance flavor (\cref{sec:LPA}). We deal with this problem in an analogous way as in the case of known values and thus take the limit on $N$ and apply our main structural theorem in \cref{sec:LimitA}. As the objective function of our problem encodes the ratio between the expected value the optimal algorithm gets and the expected maximum on the hindsight, we end up obtaining the best possible approximation guarantee for $p$-DOS with adversarial values, $\alpha(p)$, as a function of $p$, and for all values of $N$ (\cref{sec:solve}). To this end we note that the optimal algorithm, which takes the form of a sequence of thresholds, can easily be implemented for finite values of $N$ without losing in the approximation guarantee (\cref{sec:indep}). 

The value $\alpha(p)$ we obtain in \cref{sec:solve} improves upon the recent work of Kaplan et al.~\cite{KNR20} and that of Correa et al \cite{CDFSZ21}, for large values of $N$.\footnote{Since the sampling models are only equivalent in the limit.} 
More importantly, it allows to draw interesting consequences as $p$ varies. Before describing some of these let us note that by the minmax theorem $p$-DOS with adversarial values is equally hard (from an approximation guarantee perspective) if (1) the adversary chooses the $Y$ values and then the DM picks the algorithm or if (2) the adversary chooses the $Y$ values knowing the algorithm of the DM. In other words, for every value of $p$ there is a sequence $Y$ such that no algorithm for the $p$-DOS with independent sampling on this sequence can achieve an approximation better than $\alpha(p)$. Interestingly, for $p\le 1/e$ we prove that $\alpha(p)=1/(e(1-p))$. This result closes a small gap left by Kaplan et al. \cite{KNR20} in the dependent sampling model and matches the tight bound in the more restricted setting in which the values are i.i.d. samples from an unknown distribution \cite{CDFS19}. Moreover, the minmax perspective above implies that for $p\le 1/e$ the secretary problem is the hardest single selection optimal stopping problem. 

On the other end of the spectrum, as $p\to 1$, the optimal performance guarantee $\alpha(1)=\lim_{p\to 1}\alpha(p)$ equals $\alpha^*\approx 0.745$.\footnote{Note that of course our problem is ill defined if $p=1$ so the right way of thinking about $p$ close to 1 is to first fix a value $p$ and then making $N$ grow large.} This is interesting since the model admits values that are not possible to cast in the i.i.d.~prophet inequality\footnote{Consider for instance the following particular case of our model where the values are correlated. With probability $1/2$ the values are i.i.d. samples of Uniform$[0,1]$ and with probability $1/2$ they are i.i.d. samples of Uniform$[1,2]$.} 
\cite[Theorem 3.4]{KNR20} (so $\alpha(p)\le \alpha^*$), where only recently it was proved that with an amount of samples linear in $n$ one can approach $\alpha^*$.\footnote{Depending on the objective function, it is not always true that with a linear number of samples one can approximate the full information case, even in the i.i.d. model. A prominent case that has been extensively studied is revenue maximization~\cite{CR14,GHZ19}. Consider the objective of maximizing the revenue using a single price, i.e., setting a threshold (or price) $T$ in order to maximize $T$ times the probability that at least one value is above $T$. If the variables are i.i.d. and equal to $n^2/(1-p)$ w.p. $1/n^2$, and $U[0,1]$ w.p. $1-1/n^2$, a revenue of $\Omega(n)$ can be achieved in a set of $(1-p)n$ variables, taking $T=n^2/(1-p)-\varepsilon$. But if we have only access to $pn$ samples, only with probability $O(1/n^2)$ we will see a high value in both sets. And most of the time we see only realizations of $U[0,1]$, in which case we cannot differentiate the instance from only $U[0,1]$ variables, where $T>1$ gives $0$ revenue, so the most we can get is $O(1)$.} 
Indeed we can show that the approximation ratio of our algorithm, $\alpha(p)$, not only converges to 0.745 but also satisfies $\alpha(p)\ge p\cdot 0.745$, for all $p\in [0,1)$. This in particular implies that if we sample a fraction $(1-\varepsilon)$ of the values our algorithms guarantees a value that is at least $(1-\varepsilon)0.745$ times the expected maximum value of the last $\varepsilon N$ values. In other words, to guarantee an approximation factor of $\alpha^*-O(\varepsilon)$ we need $O(n/\varepsilon)$ samples, rather than the $O(n/\varepsilon^6)$ of  Rubinstein et al.~\cite{RWW20}.


Besides the extreme values of $p=0$ and $p\to 1$, we obtain the best possible guarantee for all intermediate values of $p$. An interesting special case is that of $p=1/2$, i.e., when the information set and the online set are of roughly the same size. For this special case (though with dependent sampling), Kaplan et al.~\cite{KNR20} prove that a relatively simple algorithm, achieves a performance guarantee of $1-1/e$, while the current best bound evaluates to 0.649 \cite{CDFSZ21}. Here, we prove that the optimal  algorithm for $1/2$-DOS with adversarial values has an approximation guarantee of $\alpha(1/2)\approx 0.671$, thus improving upon the state of the art. 

To make the comparison between the dependent and independent sampling models more precise we prove, in \cref{sec:indep}, that the underlying optimal approximation factors in both models differ in an additive factor of at most $O(1/\sqrt{N})$, for fixed $p<1$. In particular this says that the limit approximation factor $\alpha(p)$ applies to both settings. This connection is important since in much of our analysis we use the linear program for the dependent sampling model but then apply our results in the independent sampling model. It worth mentioning here that although very similar and essentially equivalent for large $N$, the independent sampling model is somewhat smoother than the dependent one. In particular, one can immediately define it for all values of $p\in [0,1)$ and not just those for which $pN$ is integral. Furthermore, as we prove in the paper the optimal approximation factor for the independent sampling model (and any $Y$) decreases with $N$ so that the limit bounds apply for a finite number of items. On the contrary, monotonicity on the dependent sampling model seems very challenging. 

We wrap up the paper in \cref{sec:multi} by considering versions of $p$-DOS under combinatorial constraints, in particular we consider the extension of the so called matroid secretary problem \cite{BIK07} to the case in which the DM has sampling capabilities. Using our machinery from the single selection case as a black box we are able to get a number of constant competitive algorithms for several special cases of matroids. Additionally, for general matroids, we observe that the existence of a constant competitive algorithm for $p$-DOS (for any $p$) implies the existence of a constant competitive algorithm for the matroid secretary problem, which is a notoriously hard open problem. In particular we note that if the optimal competitive ratio of $p$-DOS in this setting would converge to that in some variant of the i.i.d. case (as it does in the single selection case) then we could solve this open problem.

\section{Preliminaries \label{sec:problem}}

\subsection{$p$-DOS with known values.}

We consider the following problem, which we call $p$-sample driven optimal stopping ($p$-DOS, for short). A decision maker (DM) is given list of $N$ items with associated values $Y_1\geq \cdots\geq Y_N$. Initially each item is independently sampled with probability $p$ and conform the DM’s information set (which we denote by $H$). The remaining items, which we call online set, are presented to the DM in an online fashion in random order. We call this way of conforming the information set \textit{independent} or \textit{binomial} sampling. Although the values $Y_1\geq \cdots\geq Y_N$ are known to the DM from the beginning, upon seeing an item the DM only knows its relative ranking within the items revealed so far.\footnote{We are not assuming that all values $Y_i$ are different, but we assume that there is an arbitrary tie-breaking rule that is consistent with the relative ranks revealed and selected before the process starts.} Thus only after observing the last item the DM can certainly know which item is associated to each value. The DM has to select a single item with the goal of  maximizing its expected value. To allow comparison between different values of $N$, we think about an infinite sequence $Y$. For instances of size $N$, the values are given by  the first $N$ components of $Y$, which we denote by $Y_{[N]}$.\footnote{This problem, with non-negative values, and without the sampling phase, was introduced by Mucci \cite{M73a}.} In the unlikely event that the online set is empty (i.e., all $N$ items are sampled), we give the DM a default reward of $Y_{N+1}$, the next value in the infinite sequence $Y$.\footnote{This is not very relevant, as it does not change our optimization problem: there is no decision to be made when the online set is empty. Also, we will focus primarily on the case where $N$ is large, making this event highly unlikely.} This is also the reward obtained if the decision maker selects no item. This way, the decision maker is always better off by selecting an item. When talking about instances of $N$ items $Y_{[N]}$, we implicitly include value $Y_{N+1}$ as the default reward for the empty set. When it is clear that we are working with an instance of $N$ items, we drop the subscript $[N]$ for ease of notation.

Note that we do not assume that the values are non-negative, and the sequence may even diverge to $-\infty$. This model, as simple as it is, turns out to be quite general. Indeed, even when $p=0$, it manages to capture several problems that have been exhaustively studied in the literature, including:

\begin{itemize}
    \item\emph{Secretary problem \cite{L61}.} In this classic problem, a decision maker is presented $N$ values in an online fashion. The goal of the DM is to maximize the probability of selecting the item with the highest value. This is obtained by setting $Y_1 =1$ and $Y_i=0$ for $i\geq 2$. 
    \item \emph{$(1,K)$-secretary problem \cite{G66}.} In this variant of the standard secretary problem, the goal of the decision maker is to maximize the probability of selecting one of the top $K$ valued items. This is captured by the model by setting $Y_i=1$ for $i=1,\dots,K$ and $Y_i=0$ for $i\geq k+1$.
    \item \emph{Rank minimization problem \cite{CMRS64}.} In this problem, the goal of the decision maker is to minimize the expected rank of the selected item among the $N$ items. This is captured by setting $Y_i = -i$ for $i\geq 1$.
\end{itemize}

For any given $p$, we use $\ALG$ to refer to a specific (possibly randomized) algorithm or stopping rule. We use $\ALG(Y_{[N]})$ to denote the random variable that equals value of the the item selected by $\ALG$ on instance $Y_{[N]}$. For a given sequence $Y$ and number of items $N$, our objective is to find an algorithm that maximizes $\E(\ALG(Y_{[N]}))$, where the expectation is taken over the randomness of the process and the inner randomness of the algorithm. A consequence of our results is that (for fixed $Y$) the decision to stop should only resort on the relative ranking of an item among those that have been revealed. Thus, whenever we see an item which is ranked $\ell$ among the $i$ items seen so far, we say that that item is an $\ell$-local maximum. 



\subsection{$p$-DOS with adversarial values.}

We also study the the variant of $p$-DOS where the values $Y_j$ are chosen by an adversary and are unknown to the decision maker. In this variant, our goal is to maximize the ratio of the reward obtained by the algorithm and the expected maximum in the online set. As we are maximizing over a competitive ratio, we will restrict the adversary to select only non-negative values for the items. For instances of $N$ items, we want to maximize $\beta_{N,p}$, defined as
\begin{align}
    \beta_{N,p} = \underset{\ALG \in {\A}_N}{\sup} \underset{Y\text{ decreasing}}{\inf} \frac{\E(\ALG(Y_{[N]}))}{\E(\OPT(Y_{[N]}))},\nonumber
\end{align}
where $\A_N$ is the set of algorithms for $p$-DOS. A simple coupling argument verifies that for any $0\leq p <1$, $\beta_{N,p}$ is decreasing in $N$, so for any $p$ the worst case will be when $N$ is large. With this in mind, we wish to find the value of
\begin{align}
    \beta(p) = \underset{N\to \infty}{\lim}\beta_{N,p}.
\end{align}

\subsection{Dependent sampling} In order to obtain our results for our independent sampling problem, we study the dependent sampling variant of $p$-DOS. This model was first introduced by Kaplan et al. \cite{KNR20}. In this problem, the information set is conformed by $h=\lfloor p\cdot N \rfloor$ items with probability 1, with each item being equally likely to be sampled. 
An equivalent way to think of this problem is that the $N$ items are shuffled according to a random permutation, and the first $h$ items belong to the information set. In addition, the order of the remaining $N-h$ items is determined by the permutation. For fixed $p$, we will focus on the limit of the problem as $N\to\infty$. Formally, we study
\begin{align*}
    \alpha_{N,p}=
\sup_{\ALG\in\bar{\A}_N} \inf_{Y \text{ decreasing}} \frac{\E(\ALG(Y_{[N]}))}{\E(\OPT(Y_{[N]}))},
\end{align*}
where $\bar{\A}_N$ is the set of algorithms for the dependent sampling variant of $p$-DOS. Analogously as before, we define
\begin{align*}
    \alpha(p)=
\lim_{N\to \infty} \alpha_{N,p}
\end{align*}
Naturally, as we establish in \cref{sec:indep}, for all $0\le p <1$ we have that $\alpha(p)=\beta(p)$.

\section{Known values} \label{sec:knownY}

In this section we find the optimal algorithm for the $p$-DOS problem with known $Y$. In order to do this, we present, for any amount of items in the information set, a linear program formulation whose optimal solution maps to an optimal algorithm. We then proceed to take the limit of this linear program as $N$ goes to infinity, and reveal the structure of the limit problem. This structure allows us to rewrite the problem as that of optimizing a relatively simple real function. We conclude by showing that our approach is able to easily handle a number of classic optimal stopping problems.

\subsection{Linear programming formulation\label{sec:LPK}}

We present here a linear program formulation for our problem, inspired by Buchbinder et al. \cite{BJS14}. This linear program depends on the input instance $Y$ and we denote it by $\LP_{h,N}(Y)$. Its objective function equals the expected value of an optimal algorithm for our problem, given that the information set $H$ contains exactly $h$ items. In the linear program, variable $x_{i,\ell}$ should be interpreted as the probability that the corresponding algorithm stops at step $i$ \textit{and} the item revealed at step $i$ is ranked $\ell$ highest among the $i$ items observed so far.


\begin{align}
\left(\,\LP_{h,N}(Y)\,\right)&&\underset{x}{\max} \quad Y_{N+1}\cdot \left(1-\sum_{i=h+1}^N \sum_{\ell = 1}^i x_{i,\ell}\right)&+ \sum_{j=1}^N Y_j  \sum\limits_{i=h+1}^{N} \sum\limits_{\ell=1}^{j}\frac{ix_{i,\ell}}{N}\frac{\binom{j-1}{\ell-1}\binom{N-j}{i-\ell}}{\binom{N-1}{i-1}} \nonumber\\
&&\text{s.t.} \quad  i x_{i,\ell} + \sum\limits_{j=h+1}^{i-1}\sum\limits_{s=1}^j x_{j,s}&\leq 1 
\qquad \forall i \in [N]\setminus [h],\forall \ell \in [i], \label{const:feasibility} \\
&& \qquad x_{i,\ell}&\geq 0\qquad\forall i \in [N]\setminus [h],\forall \ell \in [i].\nonumber
\end{align}

The idea behind this linear program is that constraint \eqref{const:feasibility} forms a polyhedron rich enough to contain all relevant algorithms for the problem, and we can express the reward of the algorithm in terms of the LP variables. We call constraint \eqref{const:feasibility} the feasibility constraint. 
This linear program presents three main differences with respect to that of Buchbinder et al.~\cite{BJS14}. The first is that the objective function includes arbitrary values $Y_i$. In particular we include the additional term 
$Y_{N+1}\cdot \left(1-\sum_{i=h+1}^N \sum_{\ell = 1}^i x_{i,\ell}\right)$ which forces the algorithm to stop, since in the event of not stopping an algorithm gets $Y_{N+1}$ which is not better than having stopped in the last item. This additional term is important because values may be negative. The second is that linear program variables $x_{i,\ell}$ start at index $i=h+1$. This difference reflects the fact that the first $h$ items will conform the information set, and thus can not be selected. The third difference is that in Buchbinder et al.'s linear program, variables have the form $x_{i|\ell}$, which represent instead the probability that the algorithm selects the $i$-th item \textit{given} than the $i$-th item is ranked $\ell$ among the $i$ items seen so far. This difference does not change the linear program as there exists a bijection between the solutions given by $x_{i|\ell} = i x_{i,\ell}$.

The equivalence between solving the LP and finding an optimal algorithm is roughly as follows. Let us start by the inclusion of optimizing over algorithms in solving the LP. For any algorithm $\ALG$, given that the information set contains $h$ items, we can compute $x_{i,\ell}$: the probability that the algorithm stops at step $i$ and the $i$-th item is ranked $\ell$ among the items seen so far. As the algorithm will only see ranks, this does not depend on the values $Y_j$. These probabilities $x_{i,\ell}$ will be feasible in the polyhedron. Moreover, we can write
\begin{align}
     \PP(\ALG(Y) = Y_j) = \sum\limits_{i=h+1}^{N} \sum\limits_{\ell=1}^{i}\frac{ix_{i,\ell}}{N}\frac{\binom{j-1}{\ell-1}\binom{N-j}{i-\ell}}{\binom{N-1}{i-1}}. \nonumber
\end{align}
This way, we can express the expected reward of the algorithm as a linear function of probabilities $x_{i,\ell}$.

For the other inclusion, we see that any feasible solution $x$ can be converted into an algorithm that can be applied when the information set consists of $h$ items. We call this algorithm $\ALG_x$, and it works as follows. Let the first step be $h+1$ (representing that at the first step we have already seen $h$ items from the history set). At each step $i$, stop with probability $ix_{i,\ell}/(1-\sum_{j=1}^{i-1}\sum_{\ell=1}^jx_{j,s})$ if the current item is ranked $\ell$ among the items seen so far. The probability that $\ALG_x$ stops at the $i$-th item and the $i$-th item is ranked $\ell$ among the $i$ items seen so far is precisely $x_{i,\ell}$, which concludes the inclusion of solving the LP in finding an optimal algorithm.

Lemma \ref{lem:feasibilityLemma} formalizes the previous discussion. The proof is essentially the same as the proofs in Buchbinder et al. \cite{BJS14}, but for the sake of completeness we provide it in \cref{app:feasibility}. This result says that the optimal algorithm for sequence $Y$ with $N$ items is to observe $h$ and respond using $\ALG_x$ with $x$ being the optimal solution of $\LP_{h,N}$.

\begin{restatable}{lem}{feasibilityLemma}
\label{lem:feasibilityLemma}
Conditional on the information set containing exactly $h$ items

\begin{enumerate}
    \item For any algorithm $\ALG$, denote by $x_{i,\ell}$ the probability that $\ALG$ stops at step $i$ and the $i$-th item is ranked $\ell$ among the $i$ items seen so far. Then $x$ is feasible in $\LP_{h,N}$ and the objective function evaluated at $x$ equals the expected reward of $\ALG$.

    \item The probability that $\ALG_x$ stops at the $i$-th item and the $i$-th item is ranked $\ell$ among the $i$ items observed so far is given by $x_{i,\ell}$. The expected reward of $\ALG_x$ is equal to the objective value of $x$. \label{c}

\end{enumerate}
\label{lem:feasibility}
\end{restatable}

We have seen that in the objective function, coefficients accompanying $Y_j$ are equal to the probability that the $\ALG_x$ selects the item with value $Y_j$. The following equivalent expression of the objective function results useful for establishing our results:
\begin{align}
    \E(\ALG_x(Y)||H|=h)=&\sum_{k=1}^{N+1} Y_k \PP(\ALG_x(Y) = Y_k||H|=h) \nonumber \\
    =& \sum_{k=1}^{N} (Y_k - Y_{k+1})\PP(\ALG_x(Y)\geq Y_j||H|=h)\nonumber\\ 
    &+ Y_{N+1} \PP(\ALG_x(Y)\geq Y_{N+1}||H|=h)\nonumber\\
     =& \sum_{k=1}^{N} (Y_k - Y_{k+1})\sum_{j=1}^{k} \sum\limits_{i=h+1}^{N} \sum\limits_{\ell=1}^{j}\frac{ix_{i,\ell}}{N}\frac{\binom{j-1}{\ell-1}\binom{N-j}{i-\ell}}{\binom{N-1}{i-1}} + Y_{N+1}\nonumber\\
     =& Y_1 - \sum_{k=1}^{N}(Y_k - Y_{k+1})\left(1-\sum_{j=1}^{k} \sum\limits_{i=h+1}^{N} \sum\limits_{\ell=1}^{j}\frac{ix_{i,\ell}}{N}\frac{\binom{j-1}{\ell-1}\binom{N-j}{i-\ell}}{\binom{N-1}{i-1}} \right)\notag\\
     =& Y_1 - \sum_{k=1}^{N}(Y_k - Y_{k+1})\left(1- \sum\limits_{\ell=1}^{k}\sum\limits_{i=h+1}^{N}  x_{i,\ell}
     \sum_{j=\ell}^{k}\frac{i}{N}\frac{\binom{j-1}{\ell-1}\binom{N-j}{i-\ell}}{\binom{N-1}{i-1}} \right)
     ,\label{eq:alternativeobj}
\end{align}
where we use the fact that $\PP(\ALG_x\geq Y_{N+1}||H|=h)=1$, and that $Y_{N+1}=Y_1 - \sum_{k=1}^N (Y_k-Y_{k+1})$.


\subsection{Limit problem\label{sec:LimitK}}

For an infinite sequence $Y$, a simple coupling argument, shown in \cref{app:coupling_argument}, implies that 
\[\max_{\ALG\in\mathcal{A}_N}\E\left(\ALG\left(Y_{[N]}\right)\right)\geq \max_{\ALG\in\mathcal{A}_{N+1}} \E\left(\ALG\left(Y_{[N+1]}\right)\right).\]
This means that as $N\to\infty$ the sequence of these maxima either converges or diverges to $-\infty$. We obtain the limit of the sequence analyzing  the limit of the linear programs $\LP_{\lfloor pN\rfloor,N}$. This can be done by performing a Riemann sum analysis, which captures the cases where the limit value exists. If for $q\in L^1([p,1]\times \N)$, we define the function
\begin{align}
    F_k(q)&= \sum_{\ell=1}^k\int_p^1 q(t,\ell)\sum_{j=\ell}^k \binom{j-1}{\ell-1}(1-t)^{j-\ell}t^\ell dt, \label{eq:definition_F_k_q}
\end{align}
we can write the following limit problem, $\CLP_p$, where we have dropped the dependency on $Y$ for ease of notation.
\begin{align}
(\CLP_{p})\quad & \underset{q\in L^1([p,1]\times \N)}{\sup} \quad Y_1-\sum_{k \geq 1} (Y_k-Y_{k+1})\left(1- F_k(q)\right) \nonumber\\
\text{s.t.}\qquad & tq(t,\ell) + \int\limits_{p}^{t}\sum\limits_{s\geq1} q(\tau,s)d\tau \leq 1 
&\forall t \in [p,1],\forall \ell \geq 1 \nonumber \\
& q(t,\ell)\geq 0&\forall t \in [p,1],\forall \ell \geq 1\nonumber
\end{align}

By standard arguments (see \cref{app:convergence_to_continuous_LP}), for every $p\in [0,1)$ we can show that  the limit of $\max_{\ALG\in\mathcal{A}_N}\E\left(\ALG\left(Y_{[N]}\right)\right)$, when $N\rightarrow\infty$, exists if and only if the optimal value of $\CLP_p$ is finite, and they are equal.

\subsection{Structure of optimal solution\label{sec:structure}}

We  show that $\CLP_p$ can be restricted to solutions with a very special structure. Formally we prove the following theorem.

\begin{thm}
\label{thm:structure_sol_contLP}
For a fixed $p\in[0,1)$ and a given feasible solution $q$ for $\CLP_p$, there exists another feasible solution $q^*$ such that $F_k(q^*)\geq F_k(q)$ for all $k\geq 1$, and there is a non-decreasing sequence of numbers $\{t_i\}_{i\in \N}\subseteq [p,1]$, with $t_0=p$, that satisfies that for all $\ell\in \N,  t\in [p,1]$,
\begin{align}
    tq^*(t,\ell)
    + \int_p^t \sum_{s\geq 1}
    q^*(\tau,s) d\tau 
    &= 1,\; \text{ if } t\geq t_\ell
    \label{eq:tight_const1}\\
    q^*(t,\ell)
    &= 0, \; \text{ if } t<t_\ell.
    \label{eq:tight_const2}
\end{align}
Moreover, for all $t\in [p,1]$, we have that
\begin{align}
    q^*(t,\ell)= \begin{cases}\displaystyle\frac{T_i}{t^{i+1}} &\text{ if } t\in[t_i,t_{i+1}), \ell\leq i\\
    0 & \text{ else,}\end{cases}
    \label{eq:structure_cont_solution}
\end{align}
where $T_i=\prod_{j=1}^i t_j$.
\end{thm}
\begin{proof}
The proof is done in two steps. The first is to show that we can modify $q$ without decreasing $F_k(q)$ to obtain a solution that satisfies \cref{eq:tight_const1,eq:tight_const2}. The second is to prove that a solution satisfies \cref{eq:tight_const1,eq:tight_const2}, then it is actually as in \cref{eq:structure_cont_solution}. 

A key ingredient is to study the term accompanying $q(t,\ell)$ in $F_k(q)$. Note that the term is either $0$, if $\ell> k$, or it is  
\begin{align*}
    \sum_{j=\ell}^k \binom{j-1}{\ell-1}(1-t)^{j-\ell}t^\ell\,,
\end{align*}
if $\ell\leq k$.
The property that we will extensively use is that this term is increasing in $t$ and decreasing in $\ell$. This is implied by the fact that it corresponds to the probability that a NegativeBinomial$(\ell,t)$\footnote{The number of coin tosses necessary to obtain $\ell$ heads, if the coin lands heads with probability $t$.} random variable is at most $k$. For completeness, an arithmetic proof of this fact can be found in \cref{monotonicity}. Then, we use these facts to argue that if we take a solution that is not as in the Theorem, we can modify it without reducing the objective value.

We recursively define a sequence of solutions $(q_n)_{n\geq 0}$ as follows. We start with an arbitrary feasible solution $q_0=q$ for $\CLP_p$. If $q_{n-1}$ is a feasible solution, we have that
\begin{align*}
    q_{n-1}(t,\ell)\leq \frac{1}{t}\left(
    1-\int_p^t\sum_{s\geq 1} q_{n-1}(\tau,s)\, d\tau
    \right),
    \quad
    \forall t\in[p,1], \ell\geq 1.
\end{align*}
Note also that $\frac{1}{t}\left(
    1-\int_p^t\sum_{s\geq 1} q_{n-1}(\tau,s)\, d\tau
    \right)$
is non-negative for all $t,\ell$ so there must exist a value $t_\ell(n)\in[p,1]$ such that
\begin{align*}
    \int_p^1 q_{n-1}(t,\ell)\,dt = \int_{t_\ell(n)}^1
    \frac{1}{t}\left(
    1-\int_p^t\sum_{s\geq 1} q_{n-1}(\tau,s)\, d\tau
    \right)\, dt.
\end{align*}
Thus, we define $q_n$ as
\begin{align*}
    q_n(t,\ell)=\begin{cases}\frac{1}{t}\left(
    1-\int_p^t\sum_{s\geq 1} q_{n-1}(\tau,s)\, d\tau
    \right) & \text{if } t\geq t_\ell(n)\\
    0 & \text{if }t<t_\ell(n).
    \end{cases}
\end{align*}
Now we prove a few facts about $q_n$. First, note that for all $\ell\geq 1$, $\int_p^1 q_{n}(t,\ell) dt =\int_p^1 q_{n-1}(t,\ell) dt$. Also note that we are only moving mass to the right, and therefore,
\begin{align}
    \frac{1}{t}\left(
    1-\int_p^t\sum_{s\geq 1} q_{n-1}(\tau,s)\, d\tau
    \right) \leq 
    \frac{1}{t}\left(
    1-\int_p^t\sum_{s\geq 1} q_{n}(\tau,s)\, d\tau
    \right), \quad
    \forall t\in[p,1]. \label{eq:increasing_right_hand_feasibility}
\end{align}
This implies that if $q_{n-1}$ is feasible, $q_n$ is also feasible. Also since we are only moving mass to the right, and since the term accompanying $q(t,\ell)$ in $F_k(q)$ is increasing in $t$, necessarily $F_k(q_{n-1})\leq F_k(q_n)$ for all $k\geq 1$. Moreover, notice \cref{eq:increasing_right_hand_feasibility} also implies that $t_\ell(n)\leq t_\ell(n+1)$ for all $\ell\geq 1, n\geq 1$. Since these numbers are upper bounded by $1$, they must converge to some values $t_\ell(\infty)\in [p,1]$.

We now prove that for each $\ell\geq 1$ the sequence $(q_n(\cdot,\ell))_{n\geq 1}$ has a pointwise limit $q_\infty(\cdot,\ell)$, to which it also converges under the $L^1$ norm. Note first that if $q_0=q_1$, the sequence is constant and therefore it trivially converges. If $q_0\neq q_1$, then \cref{eq:increasing_right_hand_feasibility} for $n=1$ holds with strict inequality in some interval $[\tau_1,\tau_2]\subseteq [p,1]$, and then, if $t_\ell(1)<\tau_2$ for some $\ell\geq 1$, necessarily $t_\ell(1)<t_\ell(2)$. By evaluating the feasibility constraint in $t=1$, we have that $\sum_{s\geq 1}\int_p^1 q_0(\tau,s)d\tau\leq 1$, so there is some $s^*$ such that $\int_p^1 q_0(\tau,s^*)d\tau=\max_{s\geq 1} \int_p^1 q_0(\tau,s) d\tau$. By the definition of $t_\ell(n)$, we have that $t_{s^*}(n)=\min_{s\geq 1} t_s(n)$ for all $n\geq 1$. Then, if $q_0\neq q_1$, $0\leq t_{s^*}(1)<t_{s^*}(2)\leq t_\ell(n)$ for all $\ell\geq 1, n\geq 2$. Now, the sequence of functions
\begin{align*}
    G_n(t)= \frac{1}{t}\left(
    1-\int_p^t\sum_{s\geq 1} q_{n}(\tau,s)\, d\tau
    \right)
\end{align*}
is monotone in $n$, so it has a pointwise limit. Now, from the definition of $q_n(t,\ell)$, this also implies that $q_n(t,\ell)$ has a pointwise limit $q_\infty(t,\ell)$ when $n\rightarrow \infty$. Indeed, for $t<t_\ell(\infty)$, eventually $t<t_\ell(n)$ because $t_\ell(n)\nearrow t_\ell(\infty)$, and then $q_n(t,\ell)$ becomes $0$; and for $t\geq t_\ell(\infty)$, $q_n(t,\ell)=G_n(t)$, which has a pointwise limit. Since $t_\ell(\infty)\geq t_{s^*}(2)>0$, there is some $n_0$ such that $t_\ell(n)\geq t_{s^*}(2)/2$ for all $n\geq n_0$ and then $q_n(t,\ell)$ is dominated by the constant function equal to $2/t_{s^*}(2)$, which is integrable, so by the dominated convergence theorem, it converges to $q_\infty(t,\ell)$ in $L^1([p,1])$. This is sufficient to conclude that $F_k(q_\infty)\geq F_k(q_0)$ for all $k\geq 1$, because $F_k$ is a continuous function of $q$ and involves only the first $k$ components of $q$.

We have now that
\begin{align*}
    q_\infty(t,\ell)=\begin{cases}\frac{1}{t}\left(
    1-\int_p^t\sum_{s\geq 1} q_{\infty}(\tau,s)\, d\tau
    \right) & \text{if } t\geq t_\ell(\infty)\\
    0 & \text{if }t<t_\ell(\infty).
    \end{cases}
\end{align*}
The only missing piece is the monotonicity of $t_\ell(\infty)$. In fact, they are not necessarily monotone. However, note that swapping components of $q_\infty$ does not affect its feasibility. Since the term accompanying $q(t,\ell)$ in $F_k(q)$ is decreasing in $\ell$, for all $k\geq 1$, we can swap components of $q_{\infty}$ to obtain a function $q^*$ and a sequence $(t_\ell)_{\ell\geq 1}$ such that $t_\ell \leq t_{\ell+1}$, that satisfies \cref{eq:tight_const1,eq:tight_const2}.

For the second part of the proof of the theorem we first prove that, given the sequence $(t_\ell)_{\ell\geq 1}$, \cref{eq:tight_const1,eq:tight_const2} admit a unique solution.  Then we prove that they are satisfied by the one given in \cref{eq:structure_cont_solution}. In fact, notice that for any $\ell$, in the interval $[t_\ell,t_{\ell+1}]$ all functions $q(t,\ell')$ with $\ell'\leq \ell$ are equal, and the rest are $0$. Thus, denoting this function by $y_\ell(t)$, we can rewrite \cref{eq:tight_const1} as follows.
\begin{align}
    y'_\ell(t) = -\frac{(i+1)}{t}\cdot y_\ell(t), \quad \forall t\in[t_\ell,t_{\ell+1}].
    \label{eq:ODE_structure}
\end{align}
Again by \cref{eq:tight_const1}, we have that the function has to satisfy a continuity constraint $y_\ell(t_\ell)=\frac{1}{t_{\ell}}\left(
1-\int_p^{t_{\ell}}\sum_{s\geq 1} q^*(\tau,s)d\tau
\right)$, which depends only of previous intervals, and for $\ell=1$ it evaluates as $0$. This determines the initial value in the interval. Therefore, by the Cauchy-Lipschitz theorem, \cref{eq:tight_const1,eq:tight_const2} admit a unique solution.

We are ready now to check that the function defined by \cref{eq:structure_cont_solution} satisfies our equations. In fact, it is easy to check the continuity, by noticing that $T_i/t^{i+1}_{i+1}=T_{i+1}/t^{i+2}_{i+1}$. Replacing in \cref{eq:ODE_structure} we obtain
\begin{align*}
    -(\ell+1)\frac{T_\ell}{t^{\ell+2}}=-\frac{\ell+1}{t} \cdot \frac{T_\ell}{t^{\ell+1}} \quad \forall t\in[t_\ell,t_{\ell+1}],
\end{align*}
which clearly holds.
\end{proof}
  
Let us now apply \cref{thm:structure_sol_contLP} to simplify problem $\CLP_p$. Noting that the differences $Y_{k}-Y_{k+1}$ are non-negative we can reduce the feasible set in $\CLP_p$ to just solutions satisfying equation     \eqref{eq:structure_cont_solution}. These solutions automatically satisfy the constraints in $\CLP_p$ and therefore the problem reduces to one in which the optimization is done only over the $t_i$'s for $i\geq 1$. To explicitly write this reduced problem, and slightly abusing notation, consider the functions $F_k:[0,1]^{\N}\to \mathbb{R}$ given by 
$$
F_k(t)=\sum_{j=1}^k 
    \sum\limits_{i=1}^\infty  \int\limits_{t_i}^{t_{i+1}} \sum\limits_{\ell = 1}^{j\wedge i}\frac{T_i}{\tau^{i+1}} \binom{j-1}{\ell-1}(1-\tau)^{j-\ell}\tau^{\ell} d\tau.
$$
Note that $F_k(t)=F_k(q^*)$ where $q^*$ satisfies \eqref{eq:structure_cont_solution}.  We obtain that the value of $\CLP_p$ equals that of the reduced problem:
\begin{align*}
(\RP_{p})\quad & \underset{t=(t_i)_{i\in \N}}{\sup} \quad 
Y_1-\sum_{k \geq 1} (Y_k-Y_{k+1})\left(1- F_k(t)\right)\\
& \text{s.t.}\quad  p\le t_{i} \leq t_{i+1} \le 1& \forall i\geq1.
\end{align*}
Straightforward (but tedious) calculations show that $F_k(t)$ is increasing in $t_i$ for all $i>k$, and also concave in each $t_i$ (see \cref{concaveF}). Unfortunately though, these $F_k(\cdot)$ are not jointly concave. Therefore, the reduced problem $\RP_p$ is a real optimization problem, which is concave on each $t_i$.


\subsection{Finding the optimal thresholds \label{sec:thresholds}}

We now connect $\RP_p$ back to $p$-DOS with known values. Asymptotically, a solution to $\RP_p$ has a natural algorithmic interpretation. In this asymptotic version, items are arriving continuously (and in random order) in the interval $[0,1]$ and they belong to the online set if and only if they arrive after time $p$. $\RP_p$ thus says that the optimal algorithm is characterized by times $\{t_i\}_{i\in \N}$ such that it accepts any $\ell$-local maximum from $t_\ell$ onwards. Thus, it will reject everything arriving in $[p,t_1)$, then in $[t_1,t_2)$ it will only accept a value that is the best so far, in $[t_2,t_3)$ it will only accept a value that is best or second best, and so on. Note that with this interpretation the functions $F_k(t)$ exactly represent the probability that the algorithm stops with an item whose global rank is $k$ or better.

The implementation of this algorithm for $p$-DOS with known values and $N$ items is standard. After solving the corresponding $\RP_p$ and finding the implied optimal thresholds, we sample one \emph{arrival time}, i.e., a uniform random variable in $[0,1]$, for each of the $N$ items. The items corresponding to arrival times that landed in $[0,p]$ are included in the information set while the remaining form the online set.\footnote{If the information set is already sampled and contains $h$ items, then the procedure would be to sample $N-h$ arrival times uniform in $(p,1)$ for the items in the online set.} These are inspected in increasing order of their arrival times and the sequence $\{t_i\}_{i\in \N}$ dictates the stopping time as before. It is easy to see that the expected reward is at least as large and converges to the objective value of the infinite problem (\cref{prop:monotonicity_indep_sampling} and \cref{lem:convergent_prob_binom_model}). One last thing to notice is that the algorithm we just described might not stop, although this can be easily fixed by selecting the last item if no item was to be selected.

Note that this formulation $\RP_p$ already establishes a number of facts. The first interesting consequence is that, quite naturally, the optimal algorithm for $p$-DOS with known values is given by a sequence of thresholds $t_1\le t_2 \le \ldots$ so that after time $t_i$ we accept any item whose current ranking is $i$ or better. This fact was previously shown in some special cases by Mucci \cite{M73a} and Chan et al \cite{CCJ15}. Moreover, by exploiting properties of the objective function we can show how it leads to relatively simple real optimization problems that solve various classic single selection optimal stopping problems.

Note first that if only finitely many $Y$'s are different --as often happens in classic optimal stopping problems-- then $\RP_p$ is a finite dimensional real optimization problem. Indeed, let us assume $Y_1\ge\cdots \ge Y_m>Y_{m+1}=\ldots$. Thus the objective function in $\RP_{p}$ becomes $\sum_{k=1}^m (Y_k - Y_{k+1}) F_k(t) - Y_{m+1}$. Additionally, since the $F_k(t)$ are increasing in $t_i$ for $i>k$, all terms in the objective function are increasing in $t_i$ for $i>m$, so that we may set these variables to be 1. 
With this $\RP_{p}$ becomes the finite dimensional optimization problem of maximizing, over $t\in [p,1]^m$ the function 
$$\sum_{k=1}^m (Y_k - Y_{k+1}) \sum_{j=1}^k 
    \sum_{i=1}^m \sum_{\ell = 1}^{j\wedge i}T_i 
    \binom{j-1}{\ell-1}\int_{t_i}^{t_{i+1}}(1-\tau)^{j-\ell}\tau^{\ell-i-1} d\tau.$$
This problem is concave in each variable $t_i$, since it is a non-negative linear combination of concave functions.
We moreover conjecture that it has a unique local maximizer, so that we expect that gradient descent methods work for the problem.\footnote{A consequence of this discussion is that given an instance of $p$-DOS with its corresponding $Y$ one can, in time $O(1)$, find an sequence of thresholds leading to an arbitrarily close to optimal online algorithm. To see this first note that restricting to the first $K=O(1)$ terms in the sequence of $Y$'s is enough. Then we can restrict to the finite version of $\RP_p$ in which only the variables $t_1,\ldots,t_k$ are present. Now, for these variables we evaluate the objective function in all values belonging to a fine grid of $[0,1]^K$ and keep the best value found.}
In particular, this holds in the following examples that recover some classical results in optimal stopping. 
\begin{itemize}
\item \emph{Secretary problem.} Recall that the secretary problem is recovered by setting $Y_1=1$ and $Y_i=0$ for $i>1$. With this, the problem simplifies to
\begin{align}
\max_{0\le t_i < 1} \sum_{i=1}^\infty  \int_{t_i}^{t_{i+1}} \frac{T_i}{\tau^{i}} d\tau
= \max_{0\le t_1 < 1} \int_{t_1}^{1} \frac{t_1}{\tau} d\tau
= \max_{0\leq t_1 \leq 1}  -t_1\ln(t_1)\nonumber,
\end{align}
where the first equality follows since, by the monotonicity property of $F_k(t)$, $t_2,t_3,\ldots$ approach 1 in the supremum. The problem to the right is easily solved by taking first order conditions, so we recover the classic result that $t_1 = 1/e$ and that the optimal value of $1/e$.
\item \emph{(1,2)-Secretary.} 
Here, we have that $Y_1=Y_2=1$ and $Y_i=0$ for $i>2$. So the problem is 
\begin{align*}
    \underset{0\leq t_1 \leq t_2 \leq 1}{\max} t_1^2 + 2t_1(\ln(t_2/t_1)+1) - 3t_1t_2.
\end{align*}
First order conditions give that $t_1\approx 0.347$ and $t_2 = 2/3$.\footnote{$t_1$ is the only solution of equation $x - \ln(x) = 1 + \ln(3/2)$ in (0,1).} The optimal value is approximately 0.5737, which matches the bound Gusein-Zade and Chan et al. \cite{G66,CCJ15}.
\item \emph{Minimum rank.} In this problem we seek to minimize the expected rank of the selected value, which is modeled by taking $Y_k = -k$, so that $Y_k-Y_{k+1}=1$. Thus $\RP_p$ becomes\footnote{To interpret the expression on the left, recall that $F_k(t)$ is the probability that the algorithm stops with an item whose rank is $k$ or better. Thus the objective simply represents the negative of the expected rank.}
$$
\sup_{0\le t_{i} \le 1, \, i\ge 1} \quad  -1+\sum_{k=1}^\infty 
\left( F_k(t)-1 \right) \quad
=
\sup_{0\le t_{i} \le 1, \, i\ge 1} \quad -\sum_{i=1}^\infty T_i \frac{i}{2}\left( \frac{1}{t_{i}^{i+1}} - \frac{1}{t_{i+1}^{i+1}} \right),
$$
where the equality follows by using the identity $\sum_{j=\ell}^\infty \binom{j}{\ell}(1-t)^j = (1-t)^\ell t^{-(\ell+1)}$. 
Again the first order optimality conditions are enough to solve the problem. Indeed, they solve for $t_i = \prod_{m=i}^\infty \left(\frac{m}{m+2}\right)^{1/(m+1)}$, which evaluates for an expected rank of $ \prod_{m=1}^\infty \left(\frac{m+2}{m}\right)^{1/(m+1)} \approx 3.8695$, recovering the result of Chow et al. \cite{CMRS64}.
\end{itemize}

\subsubsection*{The case of $p>0$} Although the examples we have recovered are all for the case $p=0$ we note that our results hold for general $p$. The "right" way of taking this limit is by first normalizing the objective function value. To see this note that in  $\RP_p$ (and also in $p$-DOS with known values) the values of $Y$ can be scaled without affecting the optimization problem. Thus, for instance, if $p=0$ we could scale these values to have $Y_1=1$ (so long as $Y_1>0$). This makes sense since in this situation an optimal clairvoyant algorithm will always pick $Y_1$ so that the objective of $\RP_p$ after this normalization represents the relative performance of the best online algorithm when compared to the optimal offline algorithm. For $p>0$ the expected value of the optimal offline algorithm is given by $\sum_{i=1}^\infty Y_i p^{i-1}(1-p)$. Therefore, when all $Y$'s are non-negative, the right normalization of the objective in $\RP_p$ is to divide it by this quantity. 
This leads to measuring the performance of the algorithm as the ratio between the expectation of the selected value and the expectation of the highest eligible value (the maximum value among the items in the online set). For instance, in the case of the secretary problem, for $p>1/e$, the ratio equals $p\ln(1/p)/(1-p)$.

An important remark is that this normalization does not change the optimization problem, as the denominator in the ratio depends solely on the values of $Y_i$ and $p$. However, in the next section, we consider $p$-DOS with adversarial values and therefore the $Y_i$'s become variables selected by an adversary. In this setting, the normalization is needed to appropriately measure the competitive ratio of an algorithm.

\section{Adversarial values}\label{sec:unknownY}

Up to this point we have considered that the vector of values $Y$ is known to the decision maker from the beginning. In what follows we will relax this assumption, and instead we will let the values to be chosen by an adversary. Our objective function will thus become a competitive ratio, as suggested at the end of the previous section. Consequently, we will restrict the adversary to select a decreasing sequence of non-negative values for the items. The analysis in this section will initially rely in the dependent sampling variant, where the information set is conformed of $h$ items with probability 1, and each item has equal probability of belonging to it. This model leads to a cleaner linear program and its limit naturally coincides with that for the independent sampling variant.

We start by presenting the adversary's optimization problem and use von Neumann's minmax Theorem to derive a factor revealing LP. We take the limit of this problem as $N\to \infty $ and find that our structural results of \cref{sec:structure} also hold for this limit problem. Using this structural result we reduce the limit problem to finding an optimal sequence of optimal time thresholds $(t_i)_{i\in \mathbb{N}}$. We solve this reduced problem, putting special emphasis on values of $p$ within 0 and $1/e$, on $p=1/2$, and on the limit as $p\to1$. We close the section by connecting the dependent and independent sampling models. In particular, we show that our obtained guarantees also hold for finite $N$ in the independent sampling model, while in the dependent sampling model they hold approximately with an error $\tilde O(1/\sqrt{N})$ (for fixed $p<1$).

\subsection{Factor revealing LP \label{sec:LPA}} In this subsection we present a factor-revealing linear program, whose optimal value equals the optimal competitive ratio for instances with $N$ items and history set of size $h$. We start by stating our objective function, which consists of the competitive ratio just mentioned. The benchmark we will be comparing the performance of our algorithms will be the highest value among the items of the online set. Formally, our benchmark is the expectation of random variable $\OPT(Y_{[N]})$, defined as the highest value among the items in the online set. This way, for given integers $ 0 \leq h<N$, we want to find the largest ratio between $\E(\ALG(Y_{[N]}))$ and $\E(\OPT(Y_{[N]}))$, for all instances $Y_{[N]}$ on $N$ items.

The following lemma establishes the distribution of $\OPT(Y_{[N]})$, which will be useful for formulating $\SDLP_{h,N}$.
\begin{restatable}{lem}{distributionLemma}\label{lem:distOpt}
Consider an instance $Y_{[N]}$. Then:
\begin{equation*}
\PP(\OPT(Y_{[N]}) = Y_j ) = \begin{cases}
    \frac{N-h}{N-j+1}\prod\limits_{s=0}^{j-2}\frac{h-s}{N-s} & 1\leq j \leq h+1\\
    0 & \text{otherwise.}
    \end{cases}
\end{equation*}
\end{restatable}
We proceed to present a factor revealing linear program for $p$-DOS with adversarial values and dependent sampling. For a given $N$, let $\Y_N=\{ Y_{[N]}\in \R^{N}: Y_1\geq Y_{2} \geq \cdots \geq Y_{N}\geq 0\}$ be the set of relevant feasible values that the adversary may choose.\footnote{We say relevant because for $N$ items, only the first $N+1$ of sequence $Y$ will affect the outcome for instances of $N$ items (recall $Y_{N+1}$ is the reward obtained if the DM makes no selection). Now the online set cannot be empty, so $\OPT(Y_{[N]})$ is independent of $Y_{N+1}$. This way, setting $Y_{N+1}=0$ will always be optimal for an adversary minimizing the competitive ratio.} The problem for the adversary can be stated as follows:
\begin{align*}
    \underset{Y_{[N]} \in \Y_N}{\min}\,& \underset{x}{\max} \quad  \frac{\E(\ALG_x(Y_{[N]}))}{\E(\OPT(Y_{[N]}))}  &\nonumber\\
    &s.t.  \quad ix_{i,\ell} + \sum_{j=1}^{i-1} \sum_{s=1}^j x_{j,s} \leq 1  &\forall i\in[N]\setminus [h+1],\, \forall \ell \in [i]\\
    & \quad\quad \,\,\, x_{i,\ell} \geq 0  & \forall i\in[N]\setminus [h+1],\, \forall \ell \in [i].
\end{align*}
Since we may assume $Y_{N+1}=0$, the expression in \cref{sec:knownY} for $\E(\ALG_x(Y_{[N]}))$ becomes 
\[ \E(\ALG_x(Y_{[N]})) = \sum_{j=1}^N Y_j  \sum\limits_{i=h+1}^{N} \sum\limits_{\ell=1}^{j}\frac{ix_{i,\ell}}{N}\frac{\binom{j-1}{\ell-1}\binom{N-j}{i-\ell}}{\binom{N-1}{i-1}},\]
and for the dependent sampling variant we have $\E(\OPT(Y_{[N]})) = \sum_{j=1}^N Y_j \PP(\OPT(Y_{[N]})=Y_j)$. This problem is not linear, as the denominator of the objective function, $\E(\OPT(Y_{[N]}))$, depends on variables $Y_j$. However, note that we can arbitrarily scale $Y$ since the scaling will cancel out in the ratio $\E(\ALG_x(Y_{[N]}))/\E(\OPT(Y_{[N]}))$. Thus, without loss of generality, we can restrict the adversary to select values such that $\E(\OPT(Y_{[N]}))=1$. Now the problem is linear in both sets of decision variables, so we can use von Neumann's minmax theorem to change the order of the minimization and the maximization. We obtain the following problem:
\begin{align*}
    \underset{ \begin{subarray} \,  ix_{i,\ell} + \sum_{j=1}^{i-1}\sum_{s=1}^{j} x_{j,s}\leq 1 ,\,\forall i\in[N]\setminus [h], \ell \in [i],  \\\quad\quad\quad\quad\quad\quad\quad\,\, x\geq 0 \end{subarray}}{\max}\,& \underset{\begin{subarray}   \quad\quad\quad Y\in\Y_N \\\E(\OPT(Y_{[N]}))=1 \end{subarray}}{\min} \quad \E(\ALG_x(Y_{[N]})),\nonumber
\end{align*}
Through a stochastic dominance argument (presented in \cref{app:stochDom}) we finally derive our factor revealing linear program which we denote by $\SDLP_{h,N}$, short for ``Stochastic Dominance Linear Program'':
\begin{align}
(\SDLP_{h,N})&& \underset{x,\alpha}{\max}  \quad\alpha\nonumber\\
\text{s.t.} && i x_{i,\ell} + \sum\limits_{j=h+1}^{i-1}\sum\limits_{s=1}^j x_{j,s}&\leq 1 
&&&\forall i \in [N]\setminus [h],\forall \ell \in [i] \nonumber \\
&&\alpha - \frac{\displaystyle\sum\limits_{j=1}^k \displaystyle\sum\limits_{i=h+1}^{N} \displaystyle\sum\limits_{\ell=1}^{j}\frac{ix_{i,\ell}}{N}\frac{\binom{j-1}{\ell-1}\binom{N-j}{i-\ell}}{\binom{N-1}{i-1}}}{\sum\limits_{j=1}^k \frac{N-h}{N-j+1}\prod\limits_{s=0}^{j-2}\frac{h-s}{N-s}} &\leq 0 &&& \forall k \in [h+1] \nonumber \\
&& x_{i,\ell}&\geq 0&\qquad &&\forall i \in [N]\setminus [h],\forall \ell \in [I].\nonumber
\end{align}
The stochastic dominance argument says that for a given $x$, in the inner minimization problem we can focus our attention on instances of the form $Y_1=\cdots=Y_k=1$, $Y_{j}=0$ for $j\geq k+1$, for all $k\in[N]$ (each one of them normalized so that $\E(\OPT(Y))=1$).\footnote{Perhaps the easiest way to see this is that every feasible instance for the adversary is a convex combination of these instances.}

The first step is to see the second set of constraints as stochastic dominance constraints of the form
\[  \alpha - \frac{\PP(\ALG_x(Y_{[N]}) \geq Y_j)}{\PP(\OPT(Y_{[N]}) \geq Y_j )} \leq 0 \quad \forall j \in [h+1]. \]
Consequently, if $\alpha$ is feasible we can write the inequality as $\PP(\ALG_x(Y_{[N]}) \geq Y_j) \geq \alpha \PP(\OPT(Y_{[N]}) \geq Y_j)  $, integrate both sides and obtain the same bound but for the expectations instead of the probabilities. The bound in the expectations will be tight if $\alpha$ is feasible and the stochastic dominance constraint is binding for some index $k$. To see this, consider an instance $Y^k$ with $Y_i^k =1$ for $i\leq k$ and $Y_i^k=0$ for $i>k$. This way $\E(\ALG_x(Y_{[N]}^k))=\PP(\ALG_x(Y_{[N]}^k) \geq Y_k) $ and $\E(\OPT(Y_{[N]}^k)) = \PP(\OPT(Y_{[N]}^k) \geq Y_k )$. With this analysis we conclude that the optimal value of $\SDLP_{h,N}$ equals the optimal worst case competitive ratio for the dependent sampling variant of $p$-DOS with fixed $h$ and $N$. Moreover, we can recover an optimal algorithm from its optimal solution.


\subsection{The limit problem and its solution\label{sec:LimitA}} 
Similarly as in \cref{sec:LimitK}, we obtain the limit problem of  $\SDLP_{\lfloor pN \rfloor,N}$:
\begin{align}
(\SDCLP_{p})\quad & \underset{q\in L^1([p,1]\times \N),\,\alpha\in [0,1]}{\sup} \quad \alpha \nonumber\\
\text{s.t.}\qquad & tq(t,\ell) + \int\limits_{p}^{t}\sum\limits_{s\geq1} q(\tau,s)d\tau \leq 1 
&\forall t \in [p,1],\forall \ell &\geq 1 \label{eq:SDCLP_p_feasibility} \\
&\alpha \leq \frac{ F_k(q) }{1-p^k}  & \forall k&\geq 1 \label{eq:SDCLP_p_alpha} \\
& q(t,\ell)\geq 0&\forall t \in [p,1],\forall \ell &\geq 1.\nonumber
\end{align}
Now we can directly apply \cref{thm:structure_sol_contLP} to $\SDCLP_p$. For a solution $q$, consider a solution $q^*$ as in the theorem. By definition, $q^*$ satisfies \cref{eq:SDCLP_p_feasibility}; and from the fact that $F_k(q)\leq F_k(q^*)$ for all $k\geq 1$,  $q^*$ also satisfies \cref{eq:SDCLP_p_alpha} for the same $\alpha$ as $q$. We obtain the following reduced problem analogous to $\RP_p$, by noticing that for the thresholds $(t_i)_{i\in\N}$ that correspond to $q^*$ we have that $F_k(t)=F_k(q^*)$.
\begin{align*}
    (\SDRP_p) &\sup_{t=(t_i)_{i\in\N}} \min_{k\geq 1} \frac{F_k(t)}{1-p^k} &\\
    \text{s.t. }&\quad p\leq t_i\leq t_{i+1}\leq 1 &\forall i\geq 1.
\end{align*}
Recall that we defined $\alpha(p)$ as the limit of ratios $\alpha_{N,p}$, whose values correspond to the optimal value of $\SDLP_{\floor{pN},N}$. Consequently, $\alpha(p)$ equals the optimal value of $\SDRP_p$.

\subsection{Solving for different values of \texorpdfstring{$p$}{p} \label{sec:solve}} 
We proceed to obtain values of $\alpha(p)$ for $p \in [0,1)$. We start by briefly discussing the case where $0\leq p < 1/e$ and then study the limit as $p\to1$. For intermediate values of $p$, we present (almost) matching numerical bounds. Note that $\alpha(p)$ is an increasing function, as we establish, in a more general setting, with \cref{lem:differentp} in \cref{sec:multi}. As a consequence, the limit of $\alpha(p)$ as $p$ tends to 1 is well-defined.

\subsubsection*{The case $0\leq p<1/e$} For this range of $p$, we establish that $\alpha(p)=(e(1-p))^{-1}$. This closes the gap in Kaplan et al. \cite{KNR20}, where they obtain the same upper bound but a slightly weaker lower bound.\footnote{This value of $\alpha(p)$ was essentially known in a more restricted model with i.i.d. samples from an unknown distribution \cite{CDFSZ21}.} Our upper bound, which works for any $p\in [0,1)$ is shown in \cref{lem:differentp2} on a more general setting and with a simpler analysis than the one presented in Kaplan et al \cite{KNR20}. We obtain the lower bound by evaluating $t_1=1/e$ and $t_i=1$ for $i\geq 2$ in $\SDRP_p$ (i.e., the classic secretary problem algorithm). This means that the optimal algorithm will wait until seeing in total (counting both the online set and the history set) a fraction $1/e$ of $N$, and from that point on it will stop whenever we find an item whose value is larger than what has been observed so far. Our results also reveal that the hardest single selection optimal stopping problem for this range of $p$ is the secretary problem ($Y_1=1$ and the remaining values are 0). Indeed, the fact that the optimal value of $\SDRP_p$ is $(e(1-p))^{-1}$, together with von Neumann's minmax Theorem tells us that for any sequence $Y$, we can obtain a competitive ratio of at least $(e(1-p))^{-1}$. Details about this case are presented in \cref{app:p_leq_1e}.

\subsubsection*{Limit as \texorpdfstring{$p$}{p} goes to 1}
We now turn our attention to the case where $p$ is close to 1. In order to show that $\lim_{p\to 1}\alpha(p)=\alpha^*$, we will explicitly construct for each $p\in(0,1)$, a feasible solution $(\tilde{q},\tilde\alpha(p))$ for $\SDCLP_p$, and then we will show that $\lim_{p\to 1}\tilde\alpha(p) = \alpha^*$. Since for every $p$, $\tilde\alpha(p)\leq \alpha(p) \leq \alpha^*$, this would prove the result.

Fix $p\in (0,1)$ for now and recall from equation (\ref{eq:structure_cont_solution}) that we can restrict to solutions $q$ for $\SDCLP_p$ with the form
\begin{align}
    q(t,\ell)= \begin{cases} \frac{T_i}{t^{i+1}} &\text{ if } t\in[t_i,t_{i+1}], \ell\leq i\\
    0 & \text{ otherwise,}\end{cases}\label{eq:structure_cont_solution_recall}
\end{align}
where $p\leq t_1\leq t_2\leq \cdots$, and $T_i=\prod_{j=1}^i t_j$. Note that for fixed $i$ and $t\in [t_i,t_{i+1}]$, the function $f(\ell)=q(t,\ell)$ is positive and constant for $\ell \leq i$, and 0 for $\ell > i$. In particular, the function $q(t,\ell)$  is non-decreasing in $\ell$. The last property is important because of the following lemma.





\begin{lem}
\label{lem:alpha_bound_in_j}
Let $(q,\alpha)$ be a feasible solution for $\SDCLP_p$ with  $q(t,\ell)$ non-increasing in $\ell$, for all $t\in [p,1]$ and $\alpha$ maximal (i.e., such that $(q,c)$ is infeasible for any $c>\alpha$). Then we must have:
\begin{align}
    \alpha \geq \inf_{k\geq 1} \frac{1}{1-p^k} \sum_{j=1}^k \int_{p}^1  tq(t,j) dt\,.\end{align}
\end{lem}

The idea behind the construction of our explicit feasible solution for $\SDCLP_p$ is to enforce that the infimum in the lower bound of Lemma \ref{lem:alpha_bound_in_j} is attained for every $k$ simultaneously. The following lemma gives us a  characterization for all such solutions.

\begin{lem}
\label{lem:equivalent} 
Let $q$ be a function of the form  \eqref{eq:structure_cont_solution_recall} for some parameters $p=t_1\leq t_2 \leq \dots \leq 1$. The system of equations
\begin{align}
\alpha&= \frac{1}{1-p^k} \sum_{j=1}^k \int_{p}^1  tq(t,j) dt, \;\; \forall k\geq 1 \label{eq:low_bound_system}\\
    \intertext{is equivalent to}
    \alpha(1-p) &= p\ln{\frac{t_2}{p}} + p - \mu_3& \label{eq:lb_k1}\\
    \alpha(1-p)p^{k-1} &= \frac{1}{k-1}\cdot\frac{T_{k}}{t_{k}^{k-1}} - \mu_{k+1}, \qquad \forall k\geq2\label{eq:lb_kg2},\\ \text{where} \qquad 
    \mu_k &= \sum_{i=k}^\infty \frac{T_i}{t_{i}^{i-1}}\cdot\frac{1}{(i-2)(i-1)}.\nonumber
\end{align}
\end{lem}

Thanks to the previous lemma, we can restrict our search to pairs $(q,\alpha)$ satisfying \eqref{eq:structure_cont_solution_recall}, \eqref{eq:lb_k1} and \eqref{eq:lb_kg2}. The following lemma gives us one such solution.

\begin{lem}
\label{lem:construction} 
Let $p, \alpha \in (0,1)$ be arbitrary numbers. Define for each $k\geq 1$, the quantity $$\gamma_k = 1-\alpha +\alpha[kp^{k-1}-(k-1)p^k].$$ Define also the sequence of times
$t_1 = p$, $t_2 = p\exp(\alpha (1-p)^2/p)$, and inductively for $k\geq 2$ define $t_{k+1}$ as the real number satisfying
\begin{align}
\left(\frac{t_k}{t_{k+1}}\right)^{k-1}&=\frac{\gamma_k}{\gamma_{k-1}}\,.\label{eq:quotient}
\end{align}
This sequence has the following properties.
\begin{enumerate}
    \item[(i)] $(t_k)_{k\geq 1}$ is increasing.
    \item[(ii)] $\lim_{k\to \infty} t_k \leq 1$ if and only if
    \begin{align}
\ln{p}+{\frac{\alpha(1-p)^2}{p}} \leq \sum_{i=1}^{\infty}\frac{\ln(\gamma_{i+1})}{i(i+1)}\,. \label{eq:alpha_leq}\end{align}
and $\lim_{k\to \infty}t_k=1$ when equality holds in \eqref{eq:alpha_leq}.
\item[(iii)] Let $q$ be the function defined from the sequence $(t_k)_{k\geq 1}$ as in \eqref{eq:structure_cont_solution_recall}. Then $(q,\alpha)$ is feasible in $\SDCLP_p$.
\end{enumerate} 
\end{lem}

Thanks to the previous lemma, as long as \eqref{eq:alpha_leq} holds for values $p,\alpha\in (0,1)$, we obtain a solution for $\CLP_p$ of value $\alpha$. The following lemma shows that such pair of values always exists.

\begin{lem}
\label{lem:unique_alpha}
For $p\in(0,1)$, there is a unique $\tilde{\alpha}\in(0,1)$ that satisfies
\begin{align}
\ln{p}+{\frac{\tilde{\alpha}(1-p)^2}{p}} = \sum_{i=1}^{\infty}\frac{\ln( 1 - \tilde{\alpha} +\tilde{\alpha}[(i+1)p^{i}- ip^{i+1}])}{i(i+1)}\,. \label{eq:def_alpha_unique}
\end{align}
Furthermore, the map $p\mapsto \tilde{\alpha}(p)$ is continuous.
\end{lem}

We are now ready to prove the main theorem of this section. In the next statement, $\tilde{\alpha}(p)$ is the map defined in the previous lemma, $\alpha(p)$ is the optimal value of $\SDCLP_p$ and $\alpha^*$($\approx 0.745$) is the unique solution of $\int_0^1\frac{1}{y(1-\ln{y})+1/{\alpha^*}-1}dy=1.$

\begin{thm} 
For every $p\in (0,1)$, $0\leq \tilde{\alpha}(p)\leq \alpha(p) \leq \alpha^*$. Furthermore, if we define by continuity
$\tilde{\alpha}(1):=\lim_{p\to 1} \tilde\alpha(p)$, then
$\tilde{\alpha}(1)=\alpha(1)=\alpha^*$.
\label{thm:alphapto1}
\end{thm}

\subsubsection{Linear lower bound for $p$ close to 1}

For $p\in(0,1)$, we have just designed a stopping rule $\tilde{q}$ that has a competitive ratio of at least $\tilde{\alpha}(p)$. We proceed to prove that $\tilde{\alpha}(p)$ lies above the line that connects $0$ and $\alpha^*$, which has implications for problems related to $p-$DOS. Numerically, it appears that $\tilde\alpha(p)$ is actually concave, which would suffice for this purpose. Unfortunately we have not been able to prove this so we rely on the following result.

\begin{thm}
For $p\in(0,1)$, $\tilde{\alpha}(p)\geq\alpha^*p $. \label{thm:alpha_super_linear}
\end{thm}

It is worth contrasting the latter result with recent results of Correa et al.~\cite{CDFS19} and Rubinstein et al. \cite{RWW20}. They consider a more restricted model than $p$-DOS with dependent sampling, in which the decision maker sequentially observes i.i.d. values taken form a distribution $F$. Furthermore, the decision maker has, beforehand, access to a number of samples from $F$. Correa et al.~\cite{CDFS19} show that if she has access to $O(n^2/\varepsilon)$ samples then she can essentially learn $F$ and guarantee a factor of $\alpha^*-O(\varepsilon)$. Rubinstein et al.~\cite{RWW20} improve this result by showing that $O(n/\varepsilon^6)$ samples are enough to guarantee a factor of $\alpha^*-O(\varepsilon)$. Since $p$-DOS is more general than the latter setting, \cref{thm:alpha_super_linear} can be interpreted as a further improvement in this direction.\footnote{Certainly, our improvement only holds when $n$ is large compared to $1/\varepsilon$, as we are analyzing the value of the limit problem.} Indeed if we take $p=1-\varepsilon$ in \cref{thm:alpha_super_linear} the online set is of size $n=\varepsilon N$ so that our information set is of size $(1-\varepsilon)N=n(1-\varepsilon)/\varepsilon$. Thus with $O(n/\varepsilon)$ samples we guarantee a factor of $\alpha^*-O(\varepsilon)$.

\subsubsection*{Numerical bounds for $0\leq p <1$}

To close this subsection we present numerical bounds for $\SDCLP_p$ for different values of $p$. For both the upper bound we solve an optimization problem based on $\SDCLP_p$, which we call $\UBP_{p,N,k_{\max}}$. For the lower bound we solve a truncation of $\SDRP_p$, which we call $\LBP_p$. Details about these optimization problems can be found in \cref{app:numerical}.

    

\begin{figure}[t]
\includegraphics[scale=0.5]{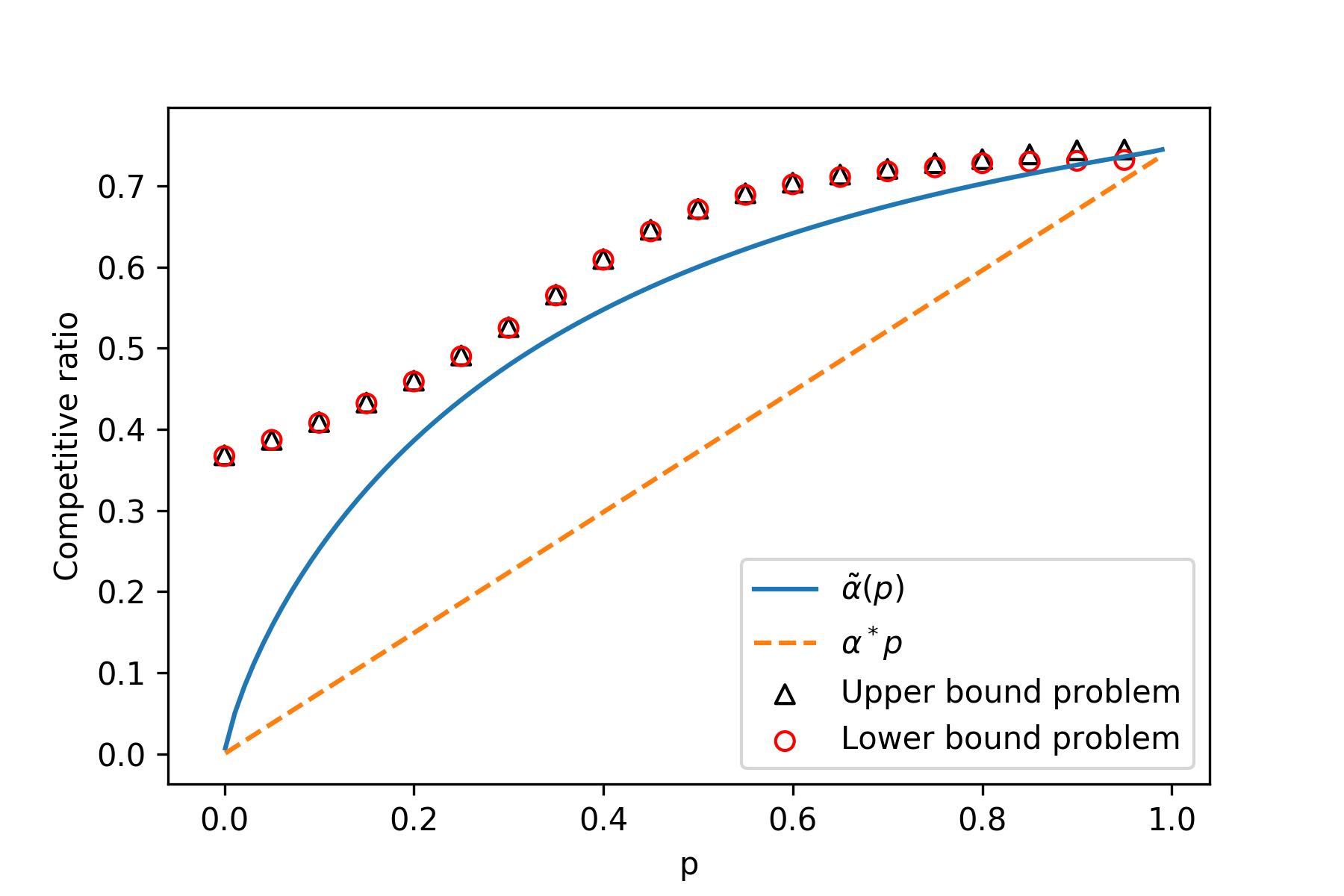}
    \caption{Plot of the numerical values of $\UBP_{p,N,k_{\max}}$ (black triangles) and $\LBP_{p,k_{\max}}$ (red circles). The blue line is $\tilde\alpha(p)$, the lower bound on $\alpha(p)$ given by \cref{thm:alphapto1}, while the orange line is $\alpha^*p$, the lower bound given by \cref{thm:alpha_super_linear}.
    \label{fig:resultados_cotas}}
\end{figure}

 In Figure \ref{fig:resultados_cotas} we plot the obtained upper and lower bounds together with the lower bound $\tilde\alpha(p)$ and the linear lower bound $\alpha^* p$. It is worth noting that $\tilde\alpha(p)$ is apparently concave but unfortunately we have not been able to prove this.

We pay special attention to the case when $p=1/2$, which corresponds to one sample for each item in the online set. In this case we obtain a lower bound of $0.671$, improving upon 0.649, the best known bound \cite{CDFSZ21}. The time thresholds for the algorithm are shown in Table  \ref{tab:p0.5}. 

\begin{table}[t]
\caption{Best found solution for $p=1/2$, rounded to the third decimal.
\label{tab:p0.5}}
\centering
\begin{tabular}{|l|l|l|l|l|l|l|l|l|l|l|}
\hline
$i$   & 1     & 2     & 3     & 4     & 5     & 6     & 7     & 8     & 9     & 10    \\ \hline
$t_i$ & 0.500 & 0.836 & 0.903 & 0.941 & 0.957 & 0.985 & 0.994 & 0.994 & 0.994 & 0.994 \\ \hline
\end{tabular}
\end{table}


\subsection{Connection between the sampling models\label{sec:indep}} 

Recall that we have defined $\alpha(p)$ and $\beta(p)$ as the limit optimal competitive ratios in the dependent and independent sampling models, respectively. So far, we have established that for any $p\in[0,1)$, $\alpha(p)$ equals $\SDRP_p$, which describes an algorithm parameterized by time thresholds $t$. We now proceed to show that $\beta(p)$ also equals to the value of $\SDRP_p$, and that this value is actually a lower bound of $\beta_{N,p}$ when $N$ is finite.

We start by relating solutions of $\SDRP_p$ with algorithms. As in \cref{sec:thresholds}, given an increasing sequence $(t_i)_{i\in\N}$, we interpret the arrival order as uniform in $[0,1]$ arrival times, and accept any $\ell$-local maximum from $t_\ell$ onwards. Let us denote this algorithm by $\ALG_t$ and its competitive ratio by
\begin{align*}
    \beta_{N,p}(t) = \inf_{Y\text{ decreasing}} \frac{\E(\ALG_t(Y_{[N]}))}{\E(\OPT(Y_{[N]}))}.
\end{align*}
Certainly, $\beta_{N,p}\geq \beta_{N,p}(t)$, for any sequence $t=(t_i)_{i\in\N}$. In the following two lemmas, we establish that in fact, for any feasible solution $t$ for $\SDRP_p$, $\beta_{N,p}(t)$ is decreasing and converges to the corresponding value of the objective function in $\SDRP_p$.

\begin{lem}
\label{prop:monotonicity_indep_sampling}
For all $N\geq 1$, $\beta_{N,p}(t) \geq \beta_{N+1,p}(t)$.
\end{lem}

\begin{lem}
\label{lem:convergent_prob_binom_model}
Fix vector $t$ of non-decreasing time thresholds. For any instance $Y$, it holds that 
\begin{align*}
    \lim_{N\rightarrow \infty} \PP(\ALG_t(Y_{[N]}) = Y_j) = \sum_{i=1}^\infty \int_{t_i}^{t_{i+1}} \sum_{\ell=1}^{j\land i} \frac{T_i}{\tau^{i}} \binom{j-1}{\ell-1} (1-\tau)^{j-\ell} \tau^{\ell-1} \,d\tau .
\end{align*}
\end{lem}

\cref{lem:convergent_prob_binom_model} implies that $\lim_{N\rightarrow \infty} \PP(\ALG_t(Y_{[N]})\geq Y_j)=F_k(t)$. This, together with the fact that the guarantee of $\ALG_t$ in instances of size $N$, as for any algorithm, is given by
\begin{align*}
    \beta_{N,p}(t)= \min_{1\leq j\leq N} \frac{\PP(\ALG_t(Y_{[N]}) \geq Y_j)}{\PP(\OPT(Y_{[N]}) \geq Y_j)} = \min_{1\leq j\leq N} \frac{\PP(\ALG_t(Y_{[N]}) \geq Y_j)}{1 - p^j},
\end{align*}
implies that the limit guarantee is the one given by $\SDRP_p$. This means that taking $t^*$ as the optimal solution of $\SDRP_p$, $\beta_{N,p}\geq \beta_{N,p}(t^*)\geq \alpha(p)$, and therefore $\beta(p)\geq \alpha(p)$.

To prove that $\beta(p)\leq \alpha(p)$, assume there is $p\in[0,1)$ such that  $\beta(p)\geq\alpha(p)+ \varepsilon$, for some $\varepsilon>0$. Fix $N$, and consider the viewpoint where each item has an independent $U[0,1]$ arrival time and is in $H$ if it arrives before $p$. Since $\beta_{N,p}\geq \beta(p)$, it is clear that there is a sufficiently small $\delta>0$ such that there is an algorithm $A$ that does not stop in $[p,p+\delta]$, that obtains at least an $(\alpha(p)+\varepsilon/2)$ fraction of the optimal offline algorithm in the independent sampling model for any instance with $N$ elements. We derive from $A$ an algorithm for the dependent sampling model in the following way: we draw $N$ independent $U[0,1]$ arrival times, randomly assign the smallest $pN$ times to the items of $H$, and the rest to the items of the online set, so that the order of arrival and order of the uniform times agree. We run $A$ as if we were in the independent sampling model, with the exception that if it selects an element of $H$, we do not stop at all and get a reward of $0$. If $N$ is large enough, at least $pN$ arrival times are smaller than $(p+\delta)N$ with probability at least $1-\varepsilon/4$, and therefore, the algorithm does not selects an item of $H$. This way, we obtain an algorithm with a guarantee of at least $\alpha(p)+\varepsilon/4$ for every large enough $N$ in the dependent sampling model, which is a contradiction. Therefore, we conclude the following theorem.
\begin{thm}
Let $t^*$ be an optimal solution for $\SDRP_p$. We have that as $N$ tends to infinity, $\beta_{N,p}(t^*)\searrow \beta(p)=\alpha(p)$.
\end{thm}
The situation for dependent sampling is a bit trickier, and it is unclear whether $\alpha_{N,p}$ is a decreasing sequence. However, we can establish that $\alpha_{N,p}$ is still close to $\alpha(p)$.
\begin{thm}
\label{thm:convergence_to_limit}
For any $p\in[0,1)$ we have that
  \begin{align*}
      \alpha_{N,p} = \alpha(p) + O\left(\frac{(\log N)^2}{(1-p)^2 \sqrt{N}}\right)\,.
  \end{align*}
\end{thm}

Summarizing the previous discussion, we obtain that for any fixed value of $N$ the guarantee obtained by our algorithm $\ALG_{t^*}$, $\alpha(p)$ applies to both sampling models. In particular, for independent sampling we have that $\beta_{N,p}\ge \alpha(p)$, while for dependent sampling we have that $\alpha_{N,p}\ge \alpha(p)-\tilde O(1/((1-p)^2\sqrt{N}))$.

\section{On Multiple-choice $p$-DOS problems \label{sec:multi}}

Until now we have focused on single selection problems. It is natural to ask whether our techniques can be used for selecting multiple items from a list subject to some combinatorial constraints, such as cardinality constraints, knapsack constraints or selecting edges that form a matching in a graph. It is possible to extend some of the linear programming machinery to tackle simple constraints such as cardinality bounds using quotas (see, \cite{BJS14, CCJ15} for particular examples), but adding more complex constraints seems difficult. Nevertheless, our resulting algorithms can be used as black boxes to obtain new results for certain multiple selection problems.

To cast the problem more precisely, we consider the following version of $p$-DOS with adversarial values. A DM is given a value $p\in [0,1)$ and an independence system $(S,\mathcal{I})$.\footnote{An independence system is a pair $(S,\mathcal{I})$, where $S$ is a finite ground set, and $\mathcal{I}$ is a family of subsets of $S$, called the independent sets of the system. The system must satisfy that the empty set is independent and that every subset of an independent set is independent.} 
An adversary assigns a non-negative weight $Y(e)$ to every element $e$ of $S$. Every element is then independently placed on the information set with probability $p$ and in the online set otherwise. As in the single selection case, the DM observes all the elements in the information set and the relative rankings of their $Y$-weights (assuming a universal tie-breaking rule). Then, the online set is revealed one by one in uniform random order. Every time an element is revealed the DM needs to irrevocably decide whether to add it or not to the solution set, while making sure that the solution set is at all times independent in $(S,\mathcal{I})$.  An algorithm for this problem is $\rho$-competitive if the expected weight of the elements in the solution set is at least $\rho$ times the expected weight of a maximum weight independent subset of the online set. An alternative way to state this is the following: for any $q$, let $S[q]$ be a random subset of $S$ obtained by adding each element of $S$ to it with probability $q$ independently. The online set of our problem behaves like $S[1-p]$. Let also $\OPT(\mathcal{I},q,Y)$ be the expectation of the maximum $Y$-weight independent subset of $S[q]$ in $\mathcal{I}$. An algorithm for $p$-DOS on $(S,\mathcal{I})$  is $\rho$-competitive if for any instance the expected $Y$-weight of its output is at least $\rho \OPT(\mathcal{I},1-p,Y)$.

Denote by $\beta_{S,\mathcal{I}}(p)$ to the maximum competitive ratio $\rho$ achievable by an algorithm for $p$-DOS on $(S,\mathcal{I})$. In general, we need to analyze entire classes of independence system at once. We tackle this in the following way. If $\mathcal{C}$ is a collection of independence systems, we define $\beta_\mathcal{C}(p)$ as the infimum over all $(S,\mathcal{I})$ in $\mathcal{C}$ of $\beta_{S,\mathcal{I}}(p)$. 
For instance, by setting $\mathcal{C}$ to be the class of all matroids of rank 1 (where $S$ can have any number of elements), we recover the single-selection $p$-DOS problem and we get   $\beta_\mathcal{C}(p)= \beta(p)=\alpha(p)$. 

When $p=0$ the $p$-DOS problem just described coincides with the generalized secretary problem by Babaioff et al. \cite{BIK07}. There is a long line of work for that problem for different independence systems, most notably for knapsack \cite{BIKK07,KTRV14}, matchings \cite{KP09,KRTV13} and many classes of matroids  (see \cite{STV18} for a recent comprehensive list). Optimal competitive ratios, again for $p=0$, are only known for the classes of uniform and transversal matroids \cite{KRTV13}, and constant competitive ratios are known for several other cases. An important open question, known as the matroid secretary conjecture, \cite{BIK07, BIKK18} is to decide whether the class $\mathcal{M}$ of all matroids admits an constant competitive algorithm (in our notation, whether $\beta_\mathcal{M}(0)>0$). The best ratio so far is parameterized on the rank $r$ of the matroid. In our notation, if $\mathcal{M}_r$ is the class of matroids of rank $r$, then  $\beta_{\mathcal{M}_r}(0)=\Omega(1/\log\log r)$~\cite{L14,FSZ18}.

The problem on general independence systems has not been studied yet for the case $p>0$, however we show in the next sections that the lower bounds on the guarantees for $p=0$ transfer directly to any $p<1$. In fact, we show that for a certain natural class of independence systems, we can further improve the guarantees for large $p$ via a reduction to the single selection case $p$-DOS problem.

\subsection{Relation among guarantees for different $p$ on a given independence system $(S,\mathcal{I})$}

The following lemma shows that for any class $\mathcal{C}$ of independence systems, $\beta_{\mathcal{C}}(p)$ is increasing in $p$. 
\begin{lem}\label{lem:differentp}
Let $p_1, p_2\in [0,1)$ with $p_1<p_2$. For any $\rho$-competitive algorithm for $p_1$-DOS on $(S,\mathcal{I})$ we can construct a $\rho$-competitive algorithm for $p_2$-DOS. Therefore, for any class $\mathcal{C}$ of independence systems,  $\beta_{\mathcal{C}}(p_1)\leq \beta_{\mathcal{C}}(p_2)$.
\end{lem}

\begin{proof} Fix $(S,\mathcal{I})$, $p_1$ and $p_2$ and let $A_1$ be any $\rho$-competitive algorithm for $p_1$-DOS on  $(S,\mathcal{I})$. Let $Y$ be any instance (that is, a map $Y\colon S\to \mathbb{R}_+$). To simplify the exposition, we assume that every $e$ in $S$ selects an arrival time $t(e)$ uniformly on $[0,1]$ at random, that the elements arrive in that order and furthermore, that the arrival times are also revealed to the algorithm $A_1$ upon arrival. Consider the algorithm $A_2$ that does the following on the instance $I$. Let $X$ be the set of elements $e$ with arrival time $t(e)<f:=(p_2-p_1)/(1-p_1)$. Note that $f\leq p_2$, so $X$ is a subset of $A_2$'s history set. The algorithm will create a new instance $Y'$, on the same system, with weight assignment $Y'(e)=0$ for all $e\in X$ and $Y'(e)=Y(e)$ for the elements outside $X$. Now, it simulates $A_1$ on $Y'$ in the following way. The simulation receives all elements of $S\setminus X$ in their arrival order as before, but all elements in $X$ will be inserted at random times uniformly. More precisely, for every $e\in X$, the algorithm selects $t'(e)$ uniformly at random on the interval $[f,1]$, and for every $e\in S\setminus X$, it sets $t'(e)=t(e)$. The simulation will consider every element that as $t'(e)\leq p_2$ as its history set and the rest as the online set (note that some elements from $X$ may fall in the history set and some may fall in the online set, but every element in $A_2$'s online set will also be in the simulation's online set), using $Y'$ as their values. Whenever the simulation accepts an element $e\in S\setminus X$, it puts $e$ on the solution set $\ALG$. The elements from $X$ that the simulation accepts are discarded. The solution set $\ALG$ is independent in $(S,\mathcal{I})$ because it is a subset of the simulation's answer. 

To analyze the algorithm, from this point onward let us condition on the set $X$. Observe that the simulated $A_1$ receives the elements of the instance given by $Y'$ in a uniform random order. Furthermore, every element $e$ is in the simulation's history set as long as $t'(e)<p_2$, which happens with probability $(p_2-f)/(1-f)=p_1$, so for all purposes, the instance behaves in the same way as in the $p_1$-DOS problem. For any realization of the times $t'$, let $\OPT_{t'}$ be an optimum $Y'$-weight set of $\{e\in S\colon t'(e)\geq p_2\}$, and and let $\OPT_t$ be an optimum $Y$-weight set of $\{e\in S\colon t(e)\geq p_2\}$. Since the elements of $X$ have $Y'$-weight 0, both $\OPT_{t'}$ and $\OPT_t$ have the same $Y$-weight. 

Now, since $A_1$ is $\rho$-competitive for $p_1$-DOS, the total $Y'$-weight of the simulation solution (which is equal to the $Y$-weight of $\ALG$) is at least $\rho$ times the expected $Y'$-weight of $\OPT_{t'}$, which in turn equals the expected $Y$-weight of $\OPT_t$. Removing the condition on $X$, we obtain that $A_2$ is $\rho$-competitive for $p_2$-DOS.

From here we deduce that $\beta_{S,\mathcal{I}}(p_1)\leq \beta_{S,\mathcal{I}}(p_2)$. Taking the infimum over all systems $(S,\mathcal{I})$ in $\mathcal{C}$ we conclude that $\beta_{\mathcal{C}}(p_1)\leq \beta_{\mathcal{C}}(p_2)$.
\end{proof}

The previous lemma has some nice consequences. If we apply it to the class $\mathcal{M}_1$ of unit rank matroids we recover that for the single-selection $p$-DOS problem $\alpha(p)$ is increasing in $p$. Furthermore, it shows that any $\rho$-competitive  algorithm for the generalized secretary problem (the $0$-DOS) on a particular class $\mathcal{C}$ can be adapted to the $p$-DOS problem without decreasing its competitive ratio. To name a few examples: for any $p$, we get a $1-\Theta(1/\sqrt{k})$-algorithm for $p$-DOS on $k$-uniform matroids (adapting Kleinberg's multiple choice secretary algorithm), we get a $1/e$-competitive algorithm for $p$-DOS on transversal matroids (adapting Kesselheim's et al.'s algorithm \cite{KRTV13}) and a $1/4$-competitive for $p$-DOS on graphical matroids (adapting Soto et al.'s algorithm \cite{STV18}), and these are the current best algorithms for all three classes.

\subsection{Better guarantees for $p$-DOS on special type of independence systems}

Babaioff et al.~\cite{BDGIT09} introduced a powerful technique to obtain algorithms for generalized secretary problems by randomly reducing them to a collection of independent parallel single-choice secretary problems. This works on any independence system satisfying a property known as the $\gamma$-partition property. \footnote{Rigorously, Babaioff et al. use $1/\gamma$ instead off $\gamma$ to define this notion, but we prefer to use values smaller than one to be consistent with the presentation of the rest of this paper.} If an independence system has the $\gamma$-partition property it is easy to create an algorithm for the associated secretary problem (the $0$-DOS case) that has competitive ratio $\gamma/e$. 

Below, we extend this construction to the $p$-DOS case using a stronger property that we call the $\gamma$-sample partition property. We will show that if a system has this particular property then one can easily obtain a $\gamma \alpha(p)$-competitive algorithm for the associated $p$-DOS problem for every $p$ (Babaioff et al.'s reduction is the special case for $p=0$). Here $\alpha(p)$ is the optimal guarantee for single-selection $p$-DOS.

\subsubsection*{Sample partition property} 
A \emph{unitary partition matroid} $(S,\mathcal{P})$ is an independence system whose ground set is partitioned into color classes $(S_0,S_1,\dots, S_m)$, where only $S_0$ may be empty, so that a set $X\subseteq S$ is independent if and only if $X$ does not contain elements from $S_0$, and $X$ contains at most 1 element restricted from each other color class.  We say that an independence system $(S,\mathcal{I})$ has the $\gamma$ sample partition property if we can (randomly) define a unitary partition matroid $(S,\mathcal{P})$ on the same ground set so that 
\begin{enumerate}
\item Every set $X$ independent in $\mathcal{P}$ is also independent in $\mathcal{I}$ 
\item For any $q\in [0,1]$, and any assignment of nonnegative weights to $S$. $$\mathbb{E}_\mathcal{P}[\OPT(\mathcal{P},q)]\geq \gamma \OPT(\mathcal{I},q).$$
\end{enumerate}

The notion of $\gamma$-partition property of Babaioff et al.~\cite{BDGIT09} is recovered if we only require property (2) to hold for $q=1$. 

\subsubsection*{Algorithm for $p$-DOS on a system $(S,\mathcal{I})$  with the $\gamma$ sample partition property.} 
On a given instance $Y$ our algorithm does the following:
\begin{itemize}
	\item Construct the random unit partition matroid $\mathcal{P}$ given by the $\gamma$ sample partition property, and let $S_1,\dots, S_m$ be the parts that have allowed size 1.
	\item Let $H=S[p]$ be the information set of $S$.
	\item Run in parallel $m$ instances of the optimal asymptotic algorithm $\ALG_{t^*}$ for single-selection $p$-DOS, one for each part $S_i$. Use $S_i\cap H$ and $S_i\setminus H$ as the history set and online set respectively on the $i$-th instance. Use the arrival times defined above on each online element. Whenever a copy of $\ALG_{t^*}$ selects an element, our algorithm also selects it. 
\end{itemize}

Let $\ALG$ be the output set of our algorithm and $Y(\ALG)$ be its weight. By construction $\ALG$ is independent in the unit partition matroid $\mathcal{P}$ and therefore also in the original independence system. So, our algorithm is correct. The following theorem gives us a bound on its competitive ratio. 

\begin{thm}\label{thm:reductiongamma} The expected weight of $\ALG$ is at least $\alpha(p)\cdot \gamma$  times $\OPT(\mathcal{I},1-p,Y)$. Therefore our algorithm for $p$-DOS on an independence system with $\gamma$ sample partition property is $\alpha(p)\cdot \gamma$-competitive, where $\alpha(p)$ is the optimal guarantee for single-selection $p$-DOS.
\end{thm}
\begin{proof}
Let us fix $\mathcal{P}$ (recall that it is allowed to be random). Since $\ALG_{t^*}$ is $\alpha(p)$-competitive for single-selection, the expected weight of $\ALG\cap S_i$ is at least $\alpha(p)$ times the expected maximum weight of $S_i\setminus H$. Summing over all $i$ we get that the expected weight of $\ALG$ (given $\mathcal{P}$) is at least $\alpha(p)$ times $\OPT(\mathcal{P}, 1-p, Y)$. Taking the expectation over $\mathcal{P}$ and using the  $\gamma$-unit partition property, we obtain 
\[\mathbb{E}_\mathcal{P}[ Y(\ALG)]\geq \alpha(p)\cdot \mathbb{E}_\mathcal{P}[ \OPT(\mathcal{P},1-p,Y)]\geq \alpha(p)\cdot \gamma \cdot \OPT(\mathcal{I},1-p,Y).\qedhere \]
\end{proof}

We can use Theorem \ref{thm:reductiongamma} above to obtain better guarantees for some classes of independence systems. First of all we observe that our  notion of $\gamma$ sample partition property, although stronger than the $\gamma$ partition property, is not really that restrictive. In fact, most (if not all) proofs that a particular system satisfy the weaker notion of $\gamma$ partition, can be adapted to the stronger version directly. 

We mentioned that this theorem can be used to get lower bounds for $\beta_\mathcal{C}(p)$ that are strictly larger than the ones available for $\beta_{\mathcal{C}}(0)$ for certain classes $\mathcal{C}$. A particularly interesting example is the class $\mathcal{G}$ of all graphic matroids.  Babaioff et al.~\cite{BDGIT09} showed that graphic matroids have the partition property for $\gamma=1/3$, and thus they got a $1/(3e)$-competitive algorithm for graphic matroids. Korula and Pál~\cite{KP09} improved this by showing that this class admits the partition property for $\gamma=1/2$, obtaining a $1/(2e)$-competitive algorithm. The current best algorithm by Soto et al.~\cite{STV18} is $1/4$-competitive and uses a different technique that does not reduce to the single-choice secretary problem. Using the monotonicity of $\beta_\mathcal{G}$, we know that $\beta_{\mathcal{G}}(p)$ is at least 1/4 for every $p$. However, it is quite simple to modify Korula and Pál's proof to show that graphic matroids have the stronger $1/2$ sample partition property. Using the algorithm given by Theorem \ref{thm:reductiongamma} we obtain that $\beta_{\mathcal{G}}(p)\geq \alpha(p)/2$. We note that $\alpha(p)/2$ grows from $1/(2e)$ when $p=0$ to $\alpha^*/2\approx 0.3725$, when $p=1$. So, for sufficiently large $p$, $\alpha(p)/2$ beats 1/4.

By adapting the proofs in \cite{BDGIT09} and \cite{S14} we get a few other classes of matroids with constant $\gamma$ sample partition property such as uniform matroids with $\gamma=1-1/e$, cographic matroids ($\gamma=1/3$), $k$-column sparse matroids ($\gamma=1/k$) and matroids of density $d$ ($\gamma=1/d$).

\subsection{Limiting problem as $p\to 1$ and consequences for the matroid secretary problem (MSP)}

In Lemma \ref{lem:differentp} we showed that for any class $\mathcal{C}$, the function $\alpha_C(p)$ is increasing, our next lemma shows that this function cannot grow extremely fast.

\begin{lem}\label{lem:differentp2}
Let $p_1, p_2\in [0,1)$ with $p_1<p_2$. For any $\rho$-competitive algorithm for $p_2$-DOS on $(S,\mathcal{I})$ we can construct a $\rho (1-p_2)/(1-p_1)$-competitive algorithm for $p_1$-DOS. As a corollary, for any class $\mathcal{C}$ of independence systems,  $\beta_{\mathcal{C}}(p_1)\geq \beta_{\mathcal{C}}(p_2)\cdot (1-p_2)/(1-p_1)$. Applying this to the single-selection problem we conclude that $\alpha(0)\geq \alpha(p)(1-p)$.
\end{lem}

\begin{proof} Fix $(S,\mathcal{I})$, $p_1$ and $p_2$ and let $A_2$ be any $\rho$-competitive algorithm for $p_2$-DOS on  $(S,\mathcal{I})$. We will use the same random arrival time interpretation of the elements of the system. Consider a new algorithm $A_1$ that on any instance $Y$ for $p_1$-DOS it simply mimics what $A_2$ would do on the same instance and arrival times (note that all the elements that $A_2$ accepts arrive after time $p_2$ so they also belong to the online set of $A_1$).  The set $\ALG$ that $A_1$ returns is independent in $(S,\mathcal{I})$. To analyze its performance, we need a simple observation. Let $S[t_1,t_2]$ denote the elements arriving between times $t_1$ and $t_2$. If $X$ is the maximum weight independent set of $S[p_1,1]$ then because of the random arrival, $X\cap S[p_2,1]$ has expected weight $Y(X)\cdot (1-p_2)/(1-p_1)$, therefore, the maximum weight independent set of $S[p_2,1]$ has at least that expected weight. Using that $A_2$ is $\rho$-competitive for $p_2$-DOS 
\[ \rho (1-p_2)/(1-p_1) \OPT(\mathcal{I},1-p_1,Y)\leq \rho \OPT(\mathcal{I},1-p_2,Y) \leq Y(\ALG).\]
From here we conclude that $A_1$ is $\rho(1-p_2)/(1-p_1)$ competitive for $p_1$-DOS, and we deduce that $\beta_{S,\mathcal{I}}(p_1)\geq (1-p_2)/(1-p_1)\beta_{S,\mathcal{I}}(p_2)$. We finish the proof taking infimum on the previous inequality over all systems $(S,\mathcal{I})$ in $\mathcal{C}$. 
\end{proof}

Recall now that for the single-selection $p$-DOS problem the limit  $\lim_{p\to 1}\alpha(p)$ coincides with the factor $\alpha^*$   associated to the single-selection i.i.d. prophet inequality with known distribution. An interesting question is whether something similar occurs for other classes of independence systems different than matroids of rank 1. For example, denote again $\mathcal{M}$ and $\mathcal{M}_r$ to denote the classes of all matroids and that of all matroids of rank $r$ respectively. Let $L=\lim_{p\to 1}\beta_{\mathcal{M}}(p)$ and $L_r=\lim_{p\to 1}\beta_{\mathcal{M}_r}(p)$ so that $L_1=\alpha^*$. It would be natural to ask whether 
there is an analog of the i.i.d. prophet inequality on matroids whose optimal competitive ratio equals $L$.

There are many candidates one could study, for example in the i.i.d.~MSP, every element of a known matroid is assigned independently a value from a known distribution, and the values are later revealed to the DM. Soto~\cite{S14} studied a generalization of the i.i.d. case known as the random-assignment MSP in which an adversary selects a list of non-negative values which are then randomly assigned to the elements to the matroid, which in turn is presented in random order to the DM. Another alternative is the prophet secretary model on matroids, studied by Ehsani et al.~\cite{EHKS18} in which every element from the matroid receives independently a value from a known distribution, which may  be different for every element. 

Proving that any of this problems behaves like the limit of $p$-DOS as $p\to 1$ on all matroids may be, in fact, a very difficult task. For if we are able to show that then we would have, indirectly, solved the matroid secretary conjecture. Indeed, for all the i.i.d., the random-assignment and the prophet secretary problem on matroids, constant competitive algorithms are known \cite{S14,EHKS18}, so if any of those cases holds then $L>0$. However, since $L=\lim_{p\to 1}\beta_{\mathcal{M}}(p)$, then there exists a sufficiently small  $\varepsilon>0$, so that $\beta_{\mathcal{M}}(1-\varepsilon)\geq L/2$. But then, by Lemma \ref{lem:differentp}, $\beta_{\mathcal{M}}(0)\geq \varepsilon L/2>0$, meaning that every matroid admits a constant competitive algorithm for the matroid secretary problem.

In any case, it is likely that neither the random-assignment nor the prophet secretary problem are the correct candidates, because if one restricts the former problem  to the class $\mathcal{M}_1$ we recover the classic secretary problem whose optimal competitive ratio is $1/e\neq \alpha^*$, and the latter becomes the single-selection prophet secretary problem with known distribution for which an upper bound of $0.732<\alpha^*$ is known~\cite{CSZ19}.

\newpage

\bibliographystyle{ACM-Reference-Format}

\newpage
\appendix

\section{Proofs of Section \ref{sec:knownY}}

\subsection{Proof of Lemma\ref{lem:feasibility} \label{app:feasibility}}
\begin{proof}If we condition on the information set containing exactly $h$ items, then we can interpret the process as follows. At the beginning, values $Y_j$ are shuffled according to a random permutation $\sigma$. That is, $\sigma(i)=j$ means that the $i$-th item in the permutation has value $Y_j$. The items in the information set will be the first $h$ items according to the permutation (i.e., $Y_{\sigma(1)},\dots,Y_{\sigma(h)}$). The online set will consist of the remaining items, which will be revealed according to the order of the permutation. That is, the order is $Y_{\sigma(h+1)},Y_{\sigma(h+2),\dots,Y_{\sigma(N)}}$.

For proving the first statement of the lemma, note that 
\begin{align*}
    x_{i\ell} & = \PP(\ALG \text{ stops at step }i \wedge \text{ }Y_{\sigma(i)}\text{ is }\ell-\text{local maximum}) \\
    & = \PP(\ALG \text{ stops at step }i | \text{ }Y_{\sigma(i)}\text{ is }\ell-\text{local maximum})\PP(Y_{\sigma(i)}\text{ is }\ell-\text{local maximum})\\
    & \leq \PP(\ALG \text{ does not stop before step }i | \text{ }Y_{\sigma(i)}\text{ is }\ell-\text{local maximum})\PP(Y_{\sigma(i)}\text{ is }\ell-\text{local maximum})\\
    & = \PP(\ALG \text{ does not stop before step }i )\PP(Y_{\sigma(i)}\text{ is }\ell-\text{local maximum})\\
    & = \left(1-\sum_{j=h+1}^{i-1}\sum_{s=1}^{j} x_{j,s} \right)\frac{1}{i}.
\end{align*}
Now, for any $1\leq j \leq N$, we can write
\begin{align*}
    \PP(\ALG = Y_j)& = \sum_{i=h+1}^{N}\PP(\ALG = Y_j \wedge \ALG \text{ stops at step }i)\\
    & = \sum_{i=h+1}^{N}\PP(Y_{\sigma(i)} = Y_j \wedge \ALG \text{ stops at step }i)\\
    & = \sum_{i=h+1}^{N}\PP(\ALG_x \text{ stops at step }i|Y_{\sigma(i)} = Y_j)\PP(Y_{\sigma(i)} = Y_j).
\end{align*}

Since $\sigma$ is a uniform random permutation, we have that $\PP(Y_{\sigma(i)} = Y_j) = 1/N$. For computing $\PP(\ALG \text{ stops at step }i|Y_{\sigma(i)} = Y_j)$ we rename the following events:
\begin{itemize}
    \item $A_i = \{\ALG$ stops at step $i\}$,
    \item $B_{i\ell} = \{Y_{\sigma(i)}$ is $\ell-$local maximum$\}$, and
    \item $C_{ij}=$ $\{Y_{\sigma(i)}=Y_j\}$,
\end{itemize}
and write
\begin{align*}
     \PP(A_i|C_{ij}) &= \sum\limits_{\ell=1}^{i} \PP(A_i| C_{ij} \wedge B_{i\ell})\PP(B_{i\ell}|C_{ij})
     = \sum_{\ell=1}^{i} \PP(A_i| B_{i\ell})\PP(B_{i\ell}|C_{ij})=  \sum_{\ell=1}^{i} ix_{i,\ell} \PP(B_{i\ell}|C_{ij}),
\end{align*}
where the second equality holds because $\ALG_x$ decides whether to stop at step $i$ based only on the relative order within the first $i$ items. The third equality comes from the fact that $\PP(A_i| B_{i\ell}) = \frac{\PP(A_i \wedge B_{i\ell})}{\PP(B_{i\ell})} = ix^\ALG_{i,\ell}$, where $\PP(B_{i\ell}) = 1/i$ because $\sigma$ is a uniform random permutation. 

To compute $\PP(B_{i\ell}|C_{ij})$, notice this is the probability that $Y_j$ is $\ell$-local maximum conditional on $\sigma(j)=i$. Now, this happens if out of the $j-1$ values that are larger than $Y_j$, exactly $\ell$ arrive within the first $i-1$ positions. Since, conditional on $\sigma(j)=i$, $\sigma$ is a random permutation of the other $N-1$ items, we have that
$$\PP(B_{i\ell}|C_{ij}) = \frac{\binom{j-1}{\ell-1} \binom{N-j}{i-\ell}}{\binom{N-1}{i-1}}.$$
Putting together the computed probabilities we conclude that
\begin{align}
     \PP(\ALG = Y_j) = \sum\limits_{i=h+1}^{N} \sum\limits_{\ell=1}^{i}\frac{ix_{i,\ell}}{N}\frac{\binom{j-1}{\ell-1}\binom{N-j}{i-\ell}}{\binom{N-1}{i-1}},\label{eq:appendix_eq_Exp}
\end{align}
so the first statement follows.

To prove the second statement, first notice that as $x$ satisfies the feasibility constraint, then $\ALG_x$ is well defined in the sense that $\frac{ix_{i,\ell}}{1-\sum_{j=1}^{i-1}\sum_{s=1}^jx_{j,s}}$ will always be between 0 and 1. We need to prove that the probability that $\ALG_x$ stops at step $i$ and $Y_{\sigma(i)}$ is $\ell$-local maximum is precisely $x_{i,\ell}$. This will be done by induction on $i$, defining the following events:
\begin{itemize}
    \item $A_i=\{\ALG_x$ stops at stage $i\}$,
    \item $B_{i,\ell} = \{Y_{\sigma(i)}$ is $\ell-$local maximum$\}$, and
    \item $R_i = \{\ALG_x$ reaches stage $i\}=\{\ALG_x$ does not stop in steps $h+1,\dots,i-1\}$.
\end{itemize}

The base case is $i=h+1$ and any $1\leq \ell \leq h+1$. Here, we have that $\PP(R_{h+1}) = 1$, so
\begin{align*}
    (h+1)x_{h+1,\ell} & = \PP(A_{h+1} | R_{h+1} \wedge B_{h+1,\ell} ) \\
    &=\PP(A_{h+1}|B_{h+1,\ell}) =\PP(A_{h+1}\wedge B_{h+1,\ell})/\PP(B_{h+1,\ell})\\
    & = (h+1)\PP(A_{h+1}\wedge B_{h+1,\ell})
\end{align*}
and we obtain the result by cancelling the $(h+1)$. For $i > h+1$ and $1\leq \ell \leq i$ we have that 
\begin{align*}
    \PP(A_i \wedge B_{i,\ell}) & = \PP(A_i \wedge B_{i,\ell} \wedge R_i)\\
    & = \PP(A_i | B_{i,\ell} \wedge R_i)\PP(B_{i,\ell} \wedge R_i)\\
    & = \PP(A_i | B_{i,\ell} \wedge R_i)\PP(B_{i,\ell})\PP(R_i),
\end{align*} 
where the first equality comes from the fact that $A_i$ is contained in $R_i$ and the last equality comes from the fact that $\ALG_x$ cannot use the ranking of $Y_{\sigma(i)}$ to stop in a stage before $i$. By the construction of $\ALG_x$, we have that $\PP(A_i | B_{i,\ell} \wedge R_i) = \frac{ix_{i\ell}}{1-\sum_{j=h+1}^{i-1} \sum_{s=1}^h x_{j,s}}$. As $\sigma$ is a random and uniform permutation, we have that $\PP(B_{i,\ell})=1/i$ for any $1\leq \ell \leq i$. The only thing left to conclude is computing $\PP(R_i)$. For this we compute
\begin{align*}
    \PP(R_i) & = 1 - \sum_{j=h+1}^{i-1}\PP(\text{Stop at step }j)\\
    & = 1 - \sum_{j=h+1}^{i-1}\sum_{s=1}^j\PP(\text{Stop at step }j \wedge Y_{\sigma(j)}\text{ is }\ell-\text{local maximum}) = 1 - \sum_{j=h+1}^{i-1}\sum_{s=1}^j x_{j,s}
\end{align*}
where the last equality holds because of our inductive hypothesis. It follows that the probability that $\ALG_x$ stops at step $i$ and $Y_{\sigma(i)}$ is $\ell$-local maximum is $x_{i,\ell}$. The second statement follows, as equation (\ref{eq:appendix_eq_Exp}) holds for any algorithm, in particular for $\ALG_x$.

\end{proof}

\subsection{Coupling argument for monotonicity}
\label{app:coupling_argument}

We take an algorithm $\ALG$ for $Y_{[N+1]}$ and obtain an algorithm for $Y_{[N]}$ with at least as much reward as for $Y_{[N+1]}$. Indeed, we define $\ALG'$ for $Y_{[N]}$ in the following way. We insert a dummy element with the smallest rank in a random position, and run $\ALG$ on the sequence of $N+1$ resulting elements. If $\ALG$ attempts to select the dummy element, $\ALG'$ simply does not stop and obtains a reward of $Y_{N+1}$. We couple both algorithms by taking the position of the dummy element to be the same as $Y_{N+1}$. Then, every time $\ALG$ selects an element in $Y_{[N+1]}$ greater than $Y_{N+1}$, $\ALG'$ selects the same element in $Y_{[N]}$. When $\ALG$ selects $Y_{N+1}$, $\ALG'$ does not stop, in which case the reward is defined as $Y_{N+1}$. If $\ALG$ does not stop, its reward is $Y_{N+2}\leq Y_{N+1}$. In all cases $\ALG'$ obtains more than $\ALG$. 

\subsection{Convergence of $\E(\ALG_N^*(Y))$ to $\CLP_p$}
\label{app:convergence_to_continuous_LP}

Denote by $\E(\ALG_N^*(Y))$ the expected reward of the optimal algorithm for a given sequence $Y$, and $N\geq 1$. We start by relaxing the problem. Given a value $Z\in (-\infty,Y_1)$, we consider the problem where we get a reward of $Z$ if the algorithm does not stop. This means we replace with $Z$ in the sequence $Y$ all values $Y_j<Z$. We denote this modified sequence by $Y^Z$.
We then proceed in three main steps. First, we prove that for fixed $Z$, when $p= h/N$ the difference between the optimal values of $\LP_{h,N}(Y^Z)$ and $\CLP_p(Y^Z)$\footnote{Here we make explicit the dependence of $\CLP_p$ on the sequence $Y$.} tends to $0$ when $N\rightarrow\infty$. Second, we prove that the optimal value of $\CLP_p(Y^Z)$ is a continuous function of $p$ and use a concentration bound to show that the expectation of the optimal algorithm $\E(\ALG_N^*(Y^Z))$ tends to the optimal value of $\CLP_p(Y^Z)$ when $N$ tends to $\infty$. And third, we conclude by making $Z$ tend to $\lim_{i\rightarrow\infty}Y_i$.

For the first step, notice that for any $Z>\lim_{i\rightarrow\infty} Y_i$, we only care about finitely many $Y_j$, so we can argue about the convergence of each element in the summations of the objective functions. Note also that for any $k\geq \ell$, if $i/N=t$, 
\begin{align}
    \sum_{j=\ell}^{k}\frac{i}{N}\frac{\binom{j-1}{\ell-1}\binom{N-j}{i-\ell}}{\binom{N-1}{i-1}} \underset{N\to \infty}{\longrightarrow}
    \sum_{j=\ell}^k \binom{j-1}{\ell-1}(1-t)^{j-\ell}t^\ell,
    \label{eq:convergence_of_term_in_LP}
\end{align}
simply because they represent the probabilities of drawing samples with or without replacement. Indeed, they correspond to the probability that we need to draw at most $k$ random elements from a total of $N$ to get at least $\ell$ from a given subset of $i$ elements.
Now, for an optimal solution $q$ of $\CLP_p(Y^Z)$, we define a solution for $\LP_{h,N}(Y^Z)$ given by
\begin{align*}
    x_{i,\ell}=\int_{\frac{i-1}{N}}^{\frac{i}{N}} q(t,\ell) dt.
\end{align*}
From the feasibility of $q$ one can easily show that $x$ is feasible for $\LP_{h,N}(Y^Z)$. This, together with \cref{eq:convergence_of_term_in_LP}, implies that the limit of the optimal value of $\LP_{h,N}(Y^Z)$ is at least the optimal value of $\CLP_p(Y^Z)$. For the opposite inequality, from an optimal solution $x^*$ of $\LP_{h,N}(Y^Z)$ and a given $\varepsilon>0$, define
\begin{align*}
    q(t,\ell)=\begin{cases}
    Nx^*_{i,\ell}(1-\varepsilon) &\text{if } \ell\leq i \text{ and } i=\lceil t\cdot N\rceil\\
    0&\text{otherwise}.
    \end{cases}
\end{align*}
For a certain $\varepsilon$ that tends to $0$ with $N$, $q$ is feasible for $\CLP_p(Y^Z)$. This, together with \cref{eq:convergence_of_term_in_LP} implies that the optimal value of $\CLP_p$ is at least the limit optimal value of $\LP_{h,N}(Y^Z)$.

Now, we show the optimal value of $\CLP_p(Y^Z)$ is continuous. In fact, note on the one hand it is decreasing, since we can take a solution $q$ for a given $p\in(0,1)$ and extend it to $[p',1]$ for $p'<p$ setting it equal to $0$ for $t\in [p',p]$. On the other hand, from a solution for $p'$ we can obtain a solution for $p$ by simply truncating it. Since $F_k(q)$ is continuous in $p$, if $p'$ is close to $p$, then the truncated solution is close to the solution for $p'$. Although the number of items in $H$ is random, when $N\rightarrow\infty$, $|H|/N$ converges to $p$, so the continuity of the value of $\CLP_p(Y^Z)$ implies that if we use the optimal solution of $\LP_{|H|,N}$, the expected reward converges to $\CLP_p(Y^Z)$.

Finally, when we make $Z$ tend to $\lim_{i\rightarrow \infty} Y_i$, the optimal value of $\CLP_p(Y^Z)$ tends to the optimal solution of $\CLP(Y)$, and the limit (when $N$ tends to infinity) of $\E(\ALG^*(Y^Z))$  tends to $\E(\ALG^*(Y))$, so we conclude that if they exist they must be equal.

\subsection{Monotonicity of $\sum_{j=\ell}^k \binom{j-1}{\ell-1}(1-t)^{j-\ell}t^\ell$}
\label{monotonicity}
\begin{lem}
\label{lem:monotone_in_t}
For any fixed $k\geq 1$,$\ell\leq k$, the term $\sum_{j=\ell}^k \binom{j-1}{\ell-1}(1-t)^{j-\ell}t^\ell$ as a function of $t\in [0,1]$ is increasing.
\end{lem}

\begin{proof}
The derivative of the function with respect to $t$ is
\begin{align*}
    & \sum_{j=\ell}^k \binom{j-1}{\ell-1} \bigg( \ell  (1-t)^{j-\ell}t^{\ell-1} - (j-\ell)(1-t)^{j-\ell-1}t^\ell
    \bigg)\\
    &= \sum_{j=\ell}^k \binom{j-1}{\ell-1} \bigg( \ell  (1-t) - (j-\ell)t
    \bigg)\cdot (1-t)^{j-\ell-1}t^{\ell-1}\\
    &= \sum_{j=\ell}^k\binom{j-1}{\ell-1} \bigg( j(1-t) - (j-\ell)
    \bigg)\cdot (1-t)^{j-\ell-1}t^{\ell-1} \\
    &=  t^{\ell-1} \sum_{j=\ell}^k\left(\binom{j}{\ell-1}(j-\ell+1) (1-t)^{j-\ell} - \binom{j-1}{\ell-1}(j-\ell)(1-t)^{j-\ell-1}
    \right)\\
    &= t^{\ell-1}\binom{k}{\ell-1} (k-\ell+1)(1-t)^{k-\ell} \geq 0\,,
\end{align*}
where in the second last equality we used the identity $\binom{j-1}{\ell-1}j=\binom{j}{\ell-1}(j-\ell+1)$, and in the last equality we reduced the telescopic sum.
\end{proof}

\begin{lem}
\label{lem:monotone_in_ell}
For any fixed $k\ge 1$ and $t\in[0,1]$,  the term $\sum_{j=\ell}^k \binom{j-1}{\ell-1}(1-t)^{j-\ell}t^\ell$ as a function of $\ell$ is decreasing.
\end{lem}

\begin{proof}
We want to prove that for $\ell\leq k-1$,
\begin{align}
    \sum_{j=\ell}^k \binom{j-1}{\ell-1}(1-t)^{j-\ell}t^\ell
    \geq \sum_{j=\ell+1}^k \binom{j-1}{\ell}(1-t)^{j-\ell-1}t^{\ell+1}\,.
    \label{eq:monotone_in_ell}
\end{align}
If we compare term by term in the sum (with the same value for $j$), we have that
\begin{align*}
    \frac{\binom{j-1}{\ell-1}(1-t)^{j-\ell}t^\ell}{\binom{j-1}{\ell}(1-t)^{j-\ell-1}t^{\ell+1}}
    &= \frac{\ell (1-t)}{(j-\ell)t}\\
    &= \frac{\ell-t\ell}{tj-t\ell}\,,
\end{align*}
which is larger than $1$ whenever $j\leq \ell/t$. Thus, we can safely conclude that \cref{eq:monotone_in_ell} is true when $k\leq \ell/t$.

On the other hand, we make use of the fact that for any $y\in (-1,1)$ and $\ell\in \mathbb{N}$, the identity $\sum_{j=\ell}^\infty \binom{j}{\ell} y^j = \frac{y^\ell}{(1-y)^{\ell+1}}$ holds true. From this it is easy to see that when $k$ tends to $\infty$, the term tends to $1$, so we can rewrite it as
\begin{align}
    \sum_{j=\ell}^k \binom{j-1}{\ell-1}(1-t)^{j-\ell}t^\ell
    = 1- \sum_{j=k+1}^\infty \binom{j-1}{\ell-1}(1-t)^{j-\ell}t^\ell\,.
\end{align}
Therefore, we can rewrite \cref{eq:monotone_in_ell} as
\begin{align*}
    \sum_{j=k+1}^\infty \binom{j-1}{\ell-1}(1-t)^{j-\ell}t^\ell
    \leq \sum_{j=k+1}^\infty \binom{j-1}{\ell}(1-t)^{j-\ell-1}t^{\ell+1}\,,
\end{align*}
and then, whenever $k>\ell/t$ we can conclude that the inequality is true by comparing term by term here.
\end{proof}

\subsection{Concavity of $F_k(t)$ in each  variable}
\label{concaveF}
\begin{proof}
We start by rearranging the sums in the definition of $F_k(t)$:
\begin{align*}
F_k(t)&=\sum_{j=1}^k 
    \sum\limits_{i=1}^\infty  \int_{t_i}^{t_{i+1}} \sum\limits_{\ell = 1}^{j\wedge i}\frac{T_i}{\tau^{i+1}} \binom{j-1}{\ell-1}(1-\tau)^{j-\ell}\tau^{\ell} d\tau.\\
    &=\sum_{\ell=1}^k\sum_{i=\ell}^\infty \int_{t_i}^{t_{i+1}}\frac{T_i}{\tau^{i+1}}\sum_{j=\ell}^k \binom{j-1}{\ell-1} (1-\tau)^{j-\ell} \tau^\ell \, d\tau.
\end{align*}
We now calculate the second derivative with respect to $t_s$, for some $s\geq 1$. Recall that we defined $T_i=\prod_{j=1}^i t_j$. Observe that in the sum indexed by $i$ the terms with $i<s-1$ do not depend of $t_s$, and the terms with $i>s$ are linear in $t_s$, so neither of them affect the second derivative. Thus, if we denote $H(\tau,\ell,k)= \sum_{j=\ell}^k \binom{j-1}{\ell-1} (1-\tau)^{j-\ell} \tau^\ell$, we have that
\begin{align*}
    \frac{\partial^2}{\partial t_s^2}F_k(t) ={}&
    \sum_{\ell=1}^k \frac{\partial^2}{\partial t_s^2} \left(
    \mathds{1}_{s-1\geq \ell}\int_{t_{s-1}}^{t_s} \frac{T_{s-1}}{\tau^s}
     H(\tau,\ell,k)\, d\tau + \mathds{1}_{s\geq \ell} \int_{t_s}^{t_{s+1}} \frac{T_s}{\tau^{s+1}}
    H(\tau,\ell,k) \, d\tau
    \right)\\
    ={}& \sum_{\ell=1}^k \frac{\partial}{\partial t_s}\left(
    \mathds{1}_{s-1\geq \ell} \frac{T_{s-1}}{t_s^s} H(t_s,\ell,k)
    - \mathds{1}_{s\geq \ell}
    \frac{T_s}{t_s^{s+1}} H(t_s,\ell,k)
    + \mathds{1}_{s\geq \ell}
    \int_{t_s}^{t_{s+1}} \frac{T_{s-1}}{\tau^{s+1}} H(\tau,\ell,k) d\tau
    \right).
\end{align*}
Now, notice that $\frac{T_{s-1}}{t_s^s}=\frac{T_s}{t_s^{s+1}}$, so,
\begin{align*}
    \frac{\partial^2}{\partial t_s^2}F_k(t) ={}&
    -\mathds{1}_{s\leq k} \frac{\partial}{\partial t_s} \frac{T_{s-1}}{t_s^s} H(t_s,s,k)
    +\sum_{\ell=1}^{\min\{k,s\}}
    \frac{\partial}{\partial t_s}\int_{t_s}^{t_{s+1}} \frac{T_{s-1}}{\tau^{s+1}} H(\tau,\ell,k) d\tau\\
    ={}& -\mathds{1}_{s\leq k} \frac{\partial}{\partial t_s} \frac{T_{s-1}}{t_s^s} H(t_s,s,k)
    -\sum_{\ell=1}^{\min\{k,s\}}
    \frac{T_{s-1}}{t_s^{s+1}} H(t_s,\ell,k).
\end{align*}
At this point it is already clear that for $s>k$ the second derivative is negative. So from now on we assume $s\leq k$. Let us expand $H(t_s,s,k)$ to calculate the last derivative.
\begin{align*}
    \frac{\partial^2}{\partial t_s^2}F_k(t) ={}&
    -\frac{\partial}{\partial t_s} \frac{T_{s-1}}{t_s^s} \sum_{j=s}^k
    \binom{j-1}{s-1} (1-t_s)^{j-s} t_s^s
    -\sum_{\ell=1}^s\frac{T_{s-1}}{t_s^{s+1}}H(t_s,\ell,k)\\
    ={}& T_{s-1}\left(
    \sum_{j=s+1}^k \binom{j-1}{s-1}(j-s)(1-t_s)^{j-s-1}
    -\sum_{\ell=1}^s \frac{1}{t_s^{s+1}}H(t_s,\ell,k)
    \right)\\
    ={}&
    T_{s-1}\left(
    \sum_{j=s+1}^k \binom{j-1}{s} s(1-t_s)^{j-s-1}
    -\sum_{\ell=1}^s \frac{1}{t_s^{s+1}} H(t_s,\ell,k)
    \right)\\
    ={}&
    T_{s-1}\left(
    \frac{s}{t_s^{s+1}} H(t_s,s+1,k)
    -\sum_{\ell=1}^s \frac{1}{t_s^{s+1}} H(t_s,\ell,k)
    \right)\\
    ={}& s\frac{T_{s-1}}{t_s^{s+1}}\left(
    H(t_s,s+1,k)-\sum_{\ell=1}^s\frac{1}{s} H(t_s,\ell,k)
    \right).
\end{align*}
To conclude, note that $H(\tau,\ell,k)$ is the probability that a \textsc{NegativeBinomial}$(\tau,\ell)$ is at most $k$, i.e., the probability that at most $k$ independent coin tosses are necessary to obtain $\ell$ heads, if the coin comes up head with probability $\tau$. Therefore, $H(t_s,\ell,k)\geq H(t_s,s+1,k)$ for all $\ell\leq s$, so we get that $\frac{\partial^2}{\partial t_s^2}F_k(t)\leq 0$. This implies that $F_k(t)$ is concave as a function of $t_s$, for all $s\geq 1$.
\end{proof}

\section{Proofs of Section \ref{sec:unknownY}}

\subsection{Proof of Lemma \ref{lem:distOpt}}

\begin{proof} For $OPT(Y) = Y_j$ we need that all $Y_i$ with $i<j$ belong to the history set. The first observation is that numbers smaller than $Y_{h+1}$ cannot be the optimum, because we would need the largest $h+1$ numbers to be in the history set, which has only $h$ items.

For $j\leq h$, as the construction of the history and the online sets are based on a random permutation, we can simulate it by sequentially inserting the numbers in $N$ slots of which $h$ will correspond to the history set and the remaining $N-h$ correspond to the online set. The probability that $OPT=Y_1$ is simply the probability that $Y_1$ lands on the online slots, i.e., $\frac{N-h}{N}=1-p$. For $Y_j$ with $1<j\leq h$, we need that the largest $j-1$ values appear in $H$. Conditional on the largest $s$ values are in $H$, the probability that $Y_{s+1}$ is also in $H$ is that it lands on the $h-s$ slots of $H$ remaining among the $N-s$ remaining slots: $\frac{h-s}{N-s}$. Once all the $j-1$ largest values landed on $H$, then we need $Y_j$ to land on the $N-h$ slots of $O$, which happens with probability $\frac{N-h}{N-j+1}$.

\end{proof}

\subsection{Derivation of $\SDLP_{h,N}$\label{app:stochDom}}

To establish the equivalence between both problems we show that for any $x$ feasible in the maximization problem, the optimal values of inner problems
\[(A) \underset{\begin{subarray}   \quad\quad\,\,  Y\in\Y_N \\E(\OPT(Y))=1 \end{subarray}}{\min} \quad \E(\ALG_x(Y))\nonumber\]
and 
\begin{align}
    (B)\quad  &\underset{\alpha}{\max}\quad  \alpha\nonumber\\
    &\text{s.t.} \quad  \alpha - \frac{\displaystyle\sum\limits_{j=1}^k \displaystyle\sum\limits_{i=h+1}^{N} \displaystyle\sum\limits_{\ell=1}^{j}\frac{ix_{i,\ell}}{N}\frac{\binom{j-1}{\ell-1}\binom{N-j}{i-\ell}}{\binom{N-1}{i-1}}}{\sum\limits_{j=1}^k \frac{N-h}{N-j+1}\prod\limits_{s=0}^{j-2}\frac{h-s}{N-s}} \leq 0 \quad \forall k \in [h+1].\label{const:stochDom}
\end{align}  
are equal. From Lemma \ref{lem:feasibility} we know that
\begin{align}
     \PP(\ALG_x(Y) = Y_j) = \sum\limits_{i=h+1}^{N} \sum\limits_{\ell=1}^{i}\frac{ix_{i,\ell}}{N}\frac{\binom{j-1}{\ell-1}\binom{N-j}{i-\ell}}{\binom{N-1}{i-1}}. \nonumber
\end{align}
From Lemma \ref{lem:distOpt} we know that
\begin{equation*}
\PP(\OPT(Y) = Y_j ) = \begin{cases}
    \frac{N-h}{N-j+1}\prod\limits_{s=0}^{j-2}\frac{h-s}{N-s} & 1\leq j \leq h+1\\
    0 & \text{otherwise.}
    \end{cases}
\end{equation*}
That way, constraint (\ref{const:stochDom}) can be read as
\begin{equation}
    \PP(\ALG_x(Y)\geq Y_k) \geq \alpha \PP(\OPT(Y) \geq Y_k) \quad \forall k\in[h+1].  \label{eq:sd}
\end{equation} 
If $\alpha$ is feasible, it will hold that $\E(\ALG_x(Y)) \geq \alpha \E(\OPT(Y))$ for any instance $Y$ of $N$ items. Indeed, we can integrate $\PP(\ALG_x(Y)\geq z)$ and $\PP(\OPT(Y)\geq z)$ at both sides of (\ref{eq:sd}) to obtain the bound, as both random variables can only equal values of items. Restricting the first $h+1$ items is enough, as $\PP(\OPT(Y)\geq Y_{h+1})=1$, and $\PP(\ALG_x(Y)\geq Y_k)$ is non-decreasing in $k$. In particular, if we restrict to $Y$ such that $\E(\OPT(Y))=1$, we get that $\E(\ALG_x(Y))\geq \alpha$. This holds for feasible $\alpha$, so it holds for the optimal solution $\alpha^*$ and we get the optimal value of problem $A$ is at least $\alpha^*$.

Now consider an optimal solution for problem $B$, $\alpha^*$. It must be the case that constraint (\ref{const:stochDom}) is binding for some $k^*$. Consider the following instance $Y^{k^*}$, where we set $Y_1 = \cdots = Y_{k^*} = \lambda_{k^*}$, and $Y_j=0$ if $j>k^*$. Here, $\lambda_{k^*}>0$ is such that $\E(\OPT(Y^{k^*}))=1$. We have that $k^*$ is binding, so
\[ \E(\ALG_x(Y_{k^*})) =  \frac{\E(\ALG_x(Y^{k^*}))}{\E(\OPT(Y^{k^*}))} = \frac{\lambda_{k^*} \PP(\ALG_x(Y^{k^*}) \geq Y^{k^*}) }{\lambda_{k^*} \PP(\OPT(Y^{k^*}) \geq Y^{k^*})} = \alpha^*. \]
Now, $Y^{k^*}$ is feasible in problem $(A)$, concluding that the optimal value of problem $A$ is at most $\E(\ALG_x(Y^{k^*})) = \alpha^*$. The equivalence between the two problems follows by replacing the inner problems.

\subsection{Solution of $\SDRP_p$ for $p<1/e$ \label{app:p_leq_1e}}

\proof{The upper bound follows immediately from Lemma \ref{lem:differentp2} (see also, Kaplan et al.~\cite[Theorem 3.8]{KNR20}, \cite{CDFS19}). To prove that the bound is tight we find a feasible solution of $\SDRP_p$ attaining this value. Take then $t_1 = 1/e$, $t_i=1$ for $i\geq 2$, we prove that the objective value of this solution is at least $1/(e(1-p))$


To this end first observe that the following inequalities hold for all $0\le p\le 1/e$.
$$
\int_{1/e}^1 \frac{(1-\tau)^{j-1}}{\tau}d\tau \geq \frac{1}{e^{j-1}}\ge p^{j-1}\,.
$$
Indeed the second inequality is direct. Note that the first is actually an equality for $j=1$ and $j=2$. Also for $j\ge 5$ the inequality follows since $\int_{1/e}^1 (1-\tau)^{j-1}/\tau d\tau \ge \int_{1/e}^1 (1-\tau)^{j-1}d\tau = (1-1/e)^j/j \ge 1/e^{j-1}$. Finally, for $j=3,4$ it follows from a straightforward calculation. 

Replacing our solution in $F_k(t)$ and using the previous inequalities we get
$$
   F_k(t) = \frac{1}{e}\sum_{j=1}^k \int_{1/e}^1 \frac{(1-\tau)^{j-1}}{\tau}d\tau \geq  \frac{1}{e}\sum_{j=1}^k p^{j-1}=\frac{1-p^k}{e(1-p)}\,.
$$
If we replace these values of $F_k(t)$ in the inner minimization of $\SDRP_p$, we get that all ratios equal $1/(e(1-p))$, as $1-p^k$ cancel out. We conclude that the considered solution is feasible and therefore the optimal value of $\SDRP_p$ (which is $\alpha(p)$) is at least $1/(e(1-p))$. 
\endproof}

\subsection{Proof of Lemma \ref{lem:alpha_bound_in_j}}

\begin{proof}
By the maximality of $\alpha$, 
\begin{align*}
    \alpha= \inf_{k\geq 1} \frac{1}{1-p^k} \sum_{j=1}^k \int_{p}^1 \sum_{\ell = 1}^j q(t,\ell)\binom{j-1}{\ell-1}(1-t)^{j-\ell}t^\ell dt\,.
\end{align*}
Since $q(t,\ell)$ is non-increasing in $\ell$ we can replace $q(t,\ell)$ by $q(t,j)$ in the inner sum to obtain
\begin{align}
\alpha\geq \inf_{k\geq 1} \frac{1}{1-p^k} \sum_{j=1}^k \int_{p}^1 \sum_{\ell = 1}^j q(t,j)\binom{j-1}{\ell-1}(1-t)^{j-\ell}t^\ell dt 
    &\geq \inf_{k\geq 1}
    \frac{1}{1-p^k} \sum_{j=1}^k \int_{p}^1  tq(t,j) \, dt \,,\nonumber
\end{align}
where we have used that for all $j\geq 1$ and $t\in [0,1]$, $\sum_{\ell=1}^j 
\binom{j-1}{\ell-1}(1-t)^{j-\ell}t^{\ell}=t$. 
\end{proof}

\subsection{Proof of Lemma \ref{lem:equivalent}}
\begin{proof}
Observe that \eqref{eq:low_bound_system} is equivalent to (i) $\int_{p}^1  tq (t,1) dt =  \alpha(1-p)$ and (ii) for $k\geq 2$, $\int_{p}^1  tq (t,k) dt = \alpha(1-p^k)- \alpha(1-p^{k-1})=\alpha(1-p)p^{k-1}$. So, we only need to check that the right hand side of \eqref{eq:lb_k1} and  \eqref{eq:lb_kg2} are $\int_{p}^1  tq (t,1) dt$ and $\int_{p}^1  tq (t,k) dt$ respectively.
Indeed, for $k\geq 2$:

\begin{align*}
    \int_{p}^1  tq (t,k) dt  &= \sum_{i=k}^{\infty} \int_{t_i}^{t_{i+1}}\frac{T_i}{t^i}dt
    = \sum_{i=k}^{\infty} \frac{1}{i-1} \left(\frac{T_{i}}{t_i^{i-1}} - \frac{T_{i}}{t_{i+1}^{i-1}} \right)=\sum_{i=k}^{\infty} \frac{1}{i-1} \left(\frac{T_{i}}{t_i^{i-1}} - \frac{T_{i+1}}{t_{i+1}^{i}} \right)\\
    & = \frac{1}{k-1}\cdot\frac{T_{k}}{t_{k}^{k-1}} + \sum_{i=k+1}^\infty \frac{T_i}{t_{i}^{i-1}}\left(\frac{1}{i-1}-\frac{1}{i-2} \right)= \frac{1}{k-1}\cdot\frac{T_{k}}{t_{k}^{k-1}} - \mu_{k+1}\,.
\end{align*}
Similarly, for $k=1$ we have
\begin{align*}
    \int_{p}^1 tq (t,1) dt &=\sum_{i=1}^{\infty} \int_{t_i}^{t_{i+1}} tq (t,1) dt
    = \int_{t_1}^{t_2} \frac{t_1}{t}dt + \sum_{i=2}^{\infty} \int_{t_i}^{t_{i+1}}\frac{T_i}{t^i}dt\\
    &= t_1\ln{\frac{t_2}{t_1}} + \frac{T_2}{t_2} - \mu_{3}= p\ln{\frac{t_2}{p}} + p - \mu_3\,.\qedhere
\end{align*}
\end{proof}

\subsection{Proof of Lemma \ref{lem:construction}}

\begin{proof}
We clearly have $t_1\leq t_2$. Furthermore, the denominator minus the numerator of the right hand side of \eqref{eq:quotient} is $(k-1)\alpha p^{k-2}(p-1)^2\geq 0$, implying that $t_{k+1}\geq t_k$ for all $k\geq 2$. This proves (i).

Since the sequence $(t_k)$ is increasing, it has a (possibly unbounded) limit. To compute it, we first take logarithm on both sides of \eqref{eq:quotient} and rearrange terms to obtain that for $k\geq 2$,
\begin{align*}
    \ln(t_{k+1})&=\ln(t_k)-\ln(\gamma_k^{1/(k-1)})+\ln(\gamma_{k-1}^{1/(k-1)})
    \intertext{iterating this formula we obtain}
    \ln(t_{k+1})&=\ln(t_2)-\sum_{i=1}^{k-1}\ln(\gamma_{i+1}^{1/i})+\sum_{i=1}^{k-1}\ln(\gamma_{i}^{1/i}) =\ln(t_2)-\sum_{i=1}^{k-1}\ln(\gamma_{i+1}^{1/i})+\sum_{i=0}^{k-2}\ln(\gamma_{i+1}^{1/(i+1)})\intertext{and since $\ln(\gamma_1)=\ln(1)=0$, and $\ln(t_2)=\ln p + \alpha(1-p)^2/p$, we get}
    \ln(t_{k+1})
    &=\ln p +\frac{\alpha(1-p)^2}{p}-\frac{1}{k+1}\ln(\gamma_k) -\sum_{i=1}^{k-1}\frac{\ln(\gamma_{i+1})}{i(i+1)}.
\end{align*}
Observe that $\lim_{k\to \infty}\gamma_k = 1-\alpha$. Thus, taking the limit on the previous expression we have
\begin{align*}
\lim_{k\to \infty}\ln (t_k) &=\ln p +\frac{\alpha(1-p)^2}{p} -\sum_{i=1}^{\infty}\frac{\ln(\gamma_{i+1})}{i(i+1)}
\end{align*}
Note that (ii) follows directly from here.

To finish the proof we use Lemma \ref{lem:equivalent}, and so we only need to show \eqref{eq:lb_k1} and \eqref{eq:lb_kg2}. For all $i\geq 2$, we have
\begin{align*}
\frac{T_i}{t_{i}^{i-1}}&=t_1\prod_{j=2}^{i}\frac{t_j}{t_i}=t_1\prod_{j=2}^{i}\prod_{\ell=j}^{i-1}\frac{t_\ell}{t_{\ell+1}}=p
\prod_{\ell=2}^{i-1}\frac{\gamma_\ell}{\gamma_{\ell-1}}=p\gamma_{i-1}.
\end{align*}

Therefore, using  formulas for geometric series we get that for $k\geq 3$
\begin{align*}
\mu_k&:=\sum_{i=k}^\infty \frac{T_i}{t_{i}^{i-1}}\cdot\frac{1}{(i-2)(i-1)}=p\sum_{i=k}^{\infty}\frac{1 - \alpha + \alpha[(i-1)p^{i-2} - (i-2)p^{i-1}]}{(i-1)(i-2)}\\
&=p(1-\alpha)\sum_{i=k}^\infty (\frac{1}{i-2}-\frac{1}{i-1})+\alpha \sum_{i=k}^\infty \frac{p^{i-2}}{(i-2)}-\alpha \sum_{i=k}^\infty \frac{p^{i-1}}{(i-1)}\\
&=\frac{p(1-\alpha+\alpha p^{k-2})}{k-2}.
\end{align*}

To see \eqref{eq:lb_k1}, we write
\begin{align*}
p\ln \frac{t_2}{p}+p-\mu_3 &= \alpha(1-p)^2+p-p(1-\alpha+\alpha p)=\alpha(1-p).
\end{align*}
And to get \eqref{eq:lb_kg2} we let $k\geq 2$ and write
\begin{align*}
\frac{1}{k-1} \frac{T_k}{t_k^{k-1}}-\mu_{k+1} &= \frac{p\gamma_{k-1}-p(1-\alpha+\alpha p^{k-1})}{k-1}\\
&=\frac{p(1-\alpha+\alpha[ (k-1)p^{k-2}-(k-2)p^{k-1}]-p(1-\alpha+\alpha p^{k-1})}{k-1}\\
&=p\alpha(p^{k-2}-p^{k-1}).\qedhere
\end{align*}
\end{proof}

\subsection{Proof of Lemma \ref{lem:unique_alpha}.}

\begin{proof}
Define the following functions as the left and right hand sides of the previous expression
\begin{align*}
    f(p,\alpha) &= \ln{p}+{\frac{\alpha(1-p)^2}{p}}\\
    g(p,\alpha) &= \sum_{i=1}^{\infty}\frac{\ln( 1 + \alpha(p^i(1 + i(1-p))-1 ))}{i(i+1)}\,.
\end{align*}

Both $f$ and $g$ are continuous functions of their domains. Furthermore, $f(p,\alpha)$ is increasing in $\alpha$. On the other hand, by Bernoulli inequality, $p^{-i}=(1-(1-p))^{-i}\geq 1+i(1-p)$. Therefore $p^i(1+i(1-p))-1\leq 0$. From here it is easy to see that $g(p,\alpha)$ is decreasing in $\alpha$. 
We now evaluate these two functions in $\alpha = 0$
\begin{align*}
    f(p,0) = \ln{p}<0 \qquad \text{ and} \qquad g(p,0) = 0 \,.
\end{align*}
For the case $\alpha = 1$, observe that $f(p,1)=\ln{p} + \frac{(1-p)^2}{p}$ is a 
convex function in $p\in(0,1)$ and it is minimized on $p=(\sqrt{5}-1)/2$. Therefore, there exists a universal constant $c$ such that $f(p,1)\geq c$ for all $p\in (0,1)$.

On the other hand, we have that as $\alpha$ increases, there is a vertical asymptote on some $\alpha_0\leq 1$ in which the function $g(p,\alpha)$ decreases to $-\infty$. Indeed if this was not the case, the formula for $g(p,1)$ would be well-defined, but simply replacing 1 on its expression yields
\begin{align*}
    g(p,1) &= \sum_{i=1}^{\infty}\frac{\ln{(p^i(1+i(1-p)))}}{i(i+1)}= \sum_{i=1}^{\infty}\frac{i\ln p + \ln(1+i(1-p)}{i(i+1))}\\
    &\leq \sum_{i=1}^{\infty}\frac{i\ln p + i(1-p)}{i(i+1))}=(\ln p +1-p)\sum_{i=1}^\infty \frac{1}{i+1}=-\infty.
\end{align*}
Summarizing, for every fixed value $p\in(0,1)$, the functions $f(p,\alpha)$ and $g(p,\alpha)$ are continuous, the former is increasing in $\alpha$, and the latter is decreasing in $\alpha$, and we also have that $f(p,0)<g(p,0)$ and there exists some value $\alpha'\in (0,1)$ such that $f(p,\alpha')>c>g(p,\alpha')$. By the intermediate value theorem there must be some value $\tilde{\alpha}(p)$ for which $f(p,\tilde{\alpha})=g(p,\tilde{\alpha})$, and by monotonicity and continuity of both functions, this value is unique and the map $p\mapsto \tilde{\alpha}(p)$ is continuous.
\end{proof}

\subsection{Proof of \cref{thm:alphapto1}}

\begin{proof}
By Lemmas \ref{lem:construction} and \ref{lem:unique_alpha}, we conclude that there is a feasible solution of $\SDCLP_p$ with value $\tilde{\alpha}(p)$. Therefore, $0\leq \tilde\alpha(p)\leq \alpha(p)$. Since $\alpha^*$ is the optimal performance achievable in the i.i.d.~model (see the Introduction), we also have $\alpha(p)\leq \alpha^*$. Thus, we only need to show that $\tilde{\alpha}(1)=\alpha^*$, for that, define the function
\begin{align}
    h(p,\eta)=\frac{\sum_{i=1}^{\infty}\frac{\ln( 1 - \eta +\eta[(i+1)p^{i}- ip^{i+1}])}{i(i+1)}}{\ln{p}+{\frac{\eta(1-p)^2}{p}}}\,.
\end{align}
and note that by definition of $\tilde{\alpha}(p)$, $h(p,\tilde{\alpha}(p))=1$

Let us study $h(p,\eta)$ as $p\to 1$. As both the numerator and the denominator go to 0 as $p\to 1$, we use l'H\^{o}pital's rule to find the limit
\begin{align*}
   \lim_{p\to 1} h(p,\eta)& = \lim_{p\to 1} \frac{\sum_{i=1}^{\infty} \frac{\eta(p^{i-1} - p^i )}{\eta( (i+1)p^i - ip^{i+1})+1 -\tilde{{\eta}}}}{\frac{1 }{p} + \eta\frac{-2p(1-p) - (1-p)^2}{p^2} } = \lim_{p\to 1} \frac{\sum_{i=1}^{\infty} \frac{p^{i-1} - p^i }{ (i+1)p^i - ip^{i+1}+\frac{1}{\eta} -1}}{\frac{1 }{p} + \eta\left( 1 - \frac{1}{p^2} \right) }\,.
\end{align*}

As $p\to 1$, the denominator in the last expression goes to 1. For the numerator, we will analyze the limit through a Riemann's integral analysis. For this we define $x_i=p^i$ (therefore $i=\ln{x_i}/\ln{p}$), so that intervals $(x_{i+1}, x_{i}]$ for $i\geq 1$ form a partition of the interval $(0,1]$, resulting in

\begin{align*}
    \lim_{p\to 1}\sum_{i=1}^{\infty} \frac{p^{i-1} - p^i }{ (i+1)p^i - ip^{i+1}+\frac{1}{\eta} -1} 
    &= \lim_{p\to 1}\sum_{i=1}^{\infty} \frac{x_{i-1} - x_i }{ (i+1)x_i - ix_{i}p+\frac{1}{\eta} -1}\\
    &=\lim_{p\to 1}\sum_{i=1}^{\infty} \frac{x_{i-1} - x_i }{ x_i(1 - \ln{x_i}\frac{p-1}{\ln{p}})+\frac{1}{\eta} -1}\\
    &= \int_{0}^1 \frac{1}{y(1-\ln{y}) + \frac{1}{\eta}-1}dy.
\end{align*}
In the last equality we first replaced  $\lim_{p\rightarrow 1}\frac{p-1}{\ln p}=1$ and then the limit of the Riemann sum. This can be justified by the fact that the sum is monotone in the term $\frac{p-1}{\ln p}$. So, for $p$ close enough to $1$ we can bound by replacing with $1-\varepsilon\leq \frac{p-1}{\ln p}\leq 1+\varepsilon$. Since the integral is continuous in the factor that accompanies $\ln y$, both bounds converge when $\varepsilon\rightarrow 0$. Replacing $\eta$ by $\alpha^*$ finishes the proof.
\end{proof}

\subsection{Proof of \cref{thm:alpha_super_linear}}
\begin{proof}
To prove this result we define $f(p)=\tilde{\alpha}(p)/p$. What we would like to prove is that $f(p)\geq \alpha^*$. For this, we replace $\tilde{\alpha}(p)= f(p)p$ in equation (\ref{eq:def_alpha_unique}):

\begin{align*}
\ln{p}+f(p)(1-p)^2 = \sum_{i=1}^{\infty}\frac{\ln( 1 - f(p)p +f(p)p[(i+1)p^{i}- ip^{i+1}])}{i(i+1)}\,.
\end{align*}

Note that the left-hand side of the equation is increasing in $f(p)$ and the right-hand side of the equation is decreasing in $f(p)$. Thus, to prove that $f(p)\geq \alpha^*$, we need to prove that
\begin{align*}
\ln{p}+ \alpha^*(1-p)^2 \leq  \sum_{i=1}^{\infty}\frac{\ln( 1 - \alpha^*p +\alpha^*p[(i+1)p^{i}- ip^{i+1}])}{i(i+1)}\,.
\end{align*}
By subtracting $\ln{p}$ the latter is equivalent to proving
\begin{align*}
\alpha^* \leq  \sum_{i=1}^{\infty}\frac{\ln( \frac{1}{p} - \alpha^*(1- p^i(1+i(1-p))))}{i(i+1)} +\alpha^*(2p-p^2) \,.
\end{align*}

To prove the inequality let us call its right hand side $a(p)$ and note that by definition of $\tilde{\alpha}(1)$, $a(1)=\alpha^*$, i.e., the inequality is tight for $p=1$. Therefore, to conclude we show that $a(p)$ is decreasing in $p$, so that the inequality holds for all $p\in (0,1)$. Indeed,
\begin{align*}
    \frac{d}{dp}a(p) & =  \sum_{i=1}^{\infty}\frac{-\frac{1}{p^2} +  i(i+1)\alpha^*(p^{i-1} - p^i) }{i(i+1)( \frac{1}{p} - \alpha^*( 1 - p^i(1+i(1-p)))} +2\alpha^*(1-p) \,.
\end{align*}
Letting
\begin{align*}
  b(p)= \frac{1}{p^2} \sum_{i=1}^{\infty} \frac{1}{i(i+1)( \frac{1}{p} - \alpha^*( 1 - p^i(1+i(1-p)))}  \quad
\text{ and }     \quad
c(p) = \sum_{i=1}^{\infty} \frac{p^{i-1}-p^i}{\frac{1}{p\alpha^*} -  1 + p^i(1+i(1-p))},
\end{align*}
we have $\frac{d}{dp}a(p) = 2\alpha^*(1-p) - b(p) + c(p)\,.$ Now, as 
as $\alpha^*(1 - p^i(1+i(1-p)))$ lies between $0$ and $\alpha^*<1$, we have that
\begin{align*}
    b(p) \geq& \frac{1}{p^2} \sum_{i=1}^{\infty} \frac{1}{i(i+1) \frac{1}{p}}= \frac{1}{p}\,.
\end{align*}

We now show that $c(p)\leq 1/p-\alpha^*(1-p)-\alpha^*\frac{1-p}{p}$. For this define $x_i = p^i$ and note that

\begin{align*}
    c(p) & = \frac{1}{p} \sum_{i=1}^{\infty} \frac{x_{i}-x_{i+1}}{\frac{1}{p\alpha^*} -  1 + x_{i}(1 + i(1-p))}\\
    & = \frac{1}{p} \sum_{i=1}^{\infty} \frac{x_{i}-x_{i+1}}{\frac{1}{p\alpha^*} -  1 + x_{i}\left( 1 + \ln{x_{i}}\frac{(1-p)}{\ln{p}}\right)}\\
    & \leq \frac{1}{p} \sum_{i=1}^{\infty} \frac{x_{i}-x_{i+1}}{\frac{1}{p\alpha^*} -  1 + x_{i}\left( 1 - p\ln{x_{i}}\right)}\\ 
    & \leq \frac{1}{p}\int_0^p \frac{1}{\frac{1}{p\alpha^*} -1 + y(1-p\ln{y}) }dy\\
    & = \frac{1}{p}\int_0^1 \frac{1}{\frac{1}{p\alpha^*} -1 + y(1-p\ln{y}) }dy 
    - \frac{1}{p}\int_p^1 \frac{1}{\frac{1}{p\alpha^*}-1+y(1-p\ln y)}dy\\
    &\leq \frac{1}{p}\int_0^1 \frac{1}{\frac{1}{p\alpha^*} -1 + y(1-p\ln{y}) }dy 
    -\alpha^*(1-p)
    \,.
\end{align*}
The first inequality comes from $(1-p)/\ln{p} \le -p$. The second inequality follows because $x_{i}>x_{i+1}$ and the function $((1/(p\alpha^*)-1+y(1-p\ln y)))^{-1}$ is decreasing in $y$. The last inequality comes from the fact that $1-y(1-p\ln(y))\in [0,1]$ when $y,p\in [0,1]$. Now, the integral in the last step can be rewritten as
\begin{align*}
    &\frac{1}{p}\int_0^1 \frac{1}{\frac{1}{\alpha^*}-1+y(1-\ln y)}dy -\frac{1-p}{p} \int_0^1 \frac{\frac{1}{p\alpha^*} + y\ln y}{\left(\frac{1}{\alpha^*} -1+y(1-\ln y)\right)\left(
    \frac{1}{p\alpha^*} -1 +y(1-p\ln y)\right)} dy\\
    &\leq  \frac{1}{p} - \frac{1-p}{p} \int_0^1
    \frac{1}{\frac{1}{\alpha^*}-1 +y(1-\ln y)}\cdot \frac{\frac{1}{p\alpha^*} - \frac{1}{e}}{\frac{1}{p\alpha^*}-1+y+\frac{p}{e}} dy\\
    &\leq \frac{1}{p} -\frac{1-p}{p} \int_0^1 \frac{\frac{1}{p\alpha^*} - \frac{1}{e}}{\frac{1}{p\alpha^*}-1+y+\frac{p}{e}} dy = \frac{1}{p} -\frac{1-p}{p}\left(\frac{1}{p\alpha^*}-\frac{1}{e}\right)\ln\left(\frac{\frac{1}{p\alpha^*} + \frac{p}{e}}{\frac{1}{p\alpha^*} + \frac{p}{e}-1} \right)\,.
\end{align*}
Here, the first inequality follows from the definition of $\alpha^*$ and fact that $-y\ln y\in [0,1/e]$ when $y\in(0,1)$. The second inequality comes from observing that $1/\left(\frac{1}{\alpha^*} -1 +y(1-\ln y\right)$ is decreasing, non-negative and integrates $1$, and $\left(\frac{1}{p\alpha^*} - \frac{1}{e}\right)/\left(\frac{1}{p\alpha^*}-1+y+\frac{p}{e}\right)$ is non-negative and decreasing. Finally, it can be checked numerically that if $\alpha^*\in [0.74,0.75]$, the term $\left(\frac{1}{p\alpha^*}-\frac{1}{e}\right)\ln\left(\frac{\frac{1}{p\alpha^*} + \frac{p}{e}}{\frac{1}{p\alpha^*} + \frac{p}{e}-1} \right)$ is at least $0.8$ for all $p\in (0,1)$. Then, since we know that $\alpha^*\approx 0.745$, we can conclude that $c(p)\geq \frac{1}{p} -\alpha^*(1-p)\left(1+\frac{1}{p}\right)$. Therefore,
\begin{align*}
    \frac{d}{dp}a(p) &= 2\alpha^*(1-p) - b(p) + c(p) \leq 2\alpha^*(1-p) - \frac{1}{p} + \frac{1}{p} -\alpha^*(1-p)\left(1+\frac{1}{p}\right)\leq 0\,.
\end{align*}
The result follows.
\end{proof}

\subsection{Details on numerical bounds\label{app:numerical}}
We now develop the optimization problems used for obtaining upper and lower bounds of $\alpha(p)$ when $p\in(0,1)$. For the upper bound, we construct a linear program based on $\SDCLP_p$. In this linear program we partition interval $(p,1)$ into $N(1-p)$ intervals of equal length. Inside of interval $(\frac{i-1}{N},\frac{i}{N}]$ we restrict variables $q(t,\ell)$ to be constant for every $\ell\geq 1$ and rename them $x_{i,\ell}$. We modify the feasibility constraints for making them slightly less restrictive (and equivalent as $N \to \infty$). In the minmax constraint we replace the term $(1-t)^{j-\ell}t^{\ell}$ by its upper bound
$\left(1-\frac{i-1}{N}\right)^{j-\ell}\left(\frac{i}{N}\right)^\ell$. To deal with the infinite number of variables and constraints, we introduce the parameter $k_{\max}$, which indicates that only the first $k_{\max}$ terms of the stochastic dominance constraint will be considered in the maximization. As only the first $k_{\max}$ amount of variables are considered in the objective function, we can consider only variables $x_{i,\ell}$ with $\ell \leq k_{\max}$. We call this problem $\UBP_{p,N,k_{\max}}$ (for Upper Bound Problem).
\begin{alignat}{5}
(\UBP_{p,N,k_{\max}})&& \underset{x,\alpha}{\max}  \quad\alpha\nonumber\\
\text{s.t.} && i x_{i,\ell} + \sum_{j=h+1}^{i-1}\sum_{s=1}^{k_{\max}} x_{j,s}&\leq 1 
&&&\forall i \in [N]\setminus [h],\forall \ell \in [k_{\max}] \nonumber\\
&&\alpha - \frac{\sum_{j=1}^k \sum_{i=h+1}^{N} \sum_{\ell=1}^{j} x_{i,\ell}  \frac{\binom{j-1}{\ell-1} \left( \frac{i}{N} \right)^\ell }{\left( 1 - \frac{i-1}{N} \right)^{\ell-j}}
}{1-p^k} &\leq 0 &&& \forall k \in [k_{\max}] \nonumber\\
&& x_{i,\ell}&\geq 0&\qquad &&\forall i \in [N]\setminus [h],\forall \ell \in [k_{\max}]\nonumber
\end{alignat}

For the lower bound, we numerically solve a truncated version of $\SDRP_{p}$, in which we use the parameter $k_{\max}$ to limit the amount of terms to be considered in the stochastic dominance constraint. As the solution must be a lower bound, we replace the denominator of the last term of the min-max problem by 1. This makes the objective function to be lower than $\SDRP_p$ by at most $p^{k_{\max}}$. As in the upper bound, reducing the number of stochastic dominance constraints also reduces the amount of variables to be considered, only needing to consider $t_i$ with $i\leq k_{\max}$. For simplicity, we fix $t_{k_{\max}+1}=1$ as a parameter. We call this problem $\LBP_{p,k_{\max}}$ (for Lower Bound Problem).
\begin{align*}
(\LBP_{p})\quad & \underset{t,\,\alpha\in [0,1]}{\max} \quad \alpha \\
\text{s.t.}\qquad &\alpha \leq \frac{1}{1-p^k} \sum_{j=1}^k   \sum_{i=1}^{k_{\max}}  \int_{t_i}^{t_{i+1}} \sum_{\ell = 1}^{j\wedge i}\frac{T_i}{\tau^{i+1}} \binom{j-1}{\ell-1}(1-\tau)^{j-\ell}\tau^{\ell} d\tau & \forall k\in [k_{\max}-1] \\
& \alpha \leq  \sum_{j=1}^{k_{\max}}   \sum_{i=1}^{k_{\max}}  \int_{t_i}^{t_{i+1}} \sum_{\ell = 1}^{j\wedge i}\frac{T_i}{\tau^{i+1}} \binom{j-1}{\ell-1}(1-\tau)^{j-\ell}\tau^{\ell} d\tau \\
& p\le t_{i} \leq t_{i+1}\le 1 & \forall i\in[k_{\max}]
\end{align*}

The bounds obtained for some values of $p$ are shown in Table \ref{tab:num_bounds}. Note that as $p$ gets closer to 1 we need more variables and therefore our upper and lower bounds are slightly off. This can definitely be improved by just considering more variables when solving $\UBP_{p,N,k_{\max}}$ and $\LBP_{p,k_{\max}}$ since they converge to each other.

\begin{table}
\caption{Upper and lower bounds obtained for multiples of 0.1. Parameters used for $\UBP_{p,N,k_{\max}}$ were $N=1000$ and $k_{\max}=\ln(N/(1-p))$. For $\LBP_{p,k_{\max}}$, $k_{\max}=\ln{0.001}/\ln{p}$ was used. For values of $p$ up to $1/e$ the bounds are exact and thus the difference simply comes from rounding.
\label{tab:num_bounds}}
\begin{tabular}{|l|l|l|l|l|l|l|l|l|l|}
\hline
$p$         & 0.1   & 0.2   & 0.3   & 0.4   & 0.5   & 0.6   & 0.7   & 0.8   & 0.9   \\ \hline
Lower bound & 0.408 & 0.459 & 0.525 & 0.609 & 0.671 & 0.702 & 0.718 & 0.728 & 0.730      \\ \hline
Upper bound & 0.409 & 0.460 & 0.526 & 0.610 & 0.672 & 0.704 & 0.721 & 0.733 & 0.744 \\ \hline
\end{tabular}
\end{table}

\subsection{Proof of \cref{prop:monotonicity_indep_sampling}}

\begin{proof}
For an instance $Y_{[N]}$ of size $N$, denote by $Y_{[N]}^{+0}$ the instance of size $N+1$ that results from appending a $0$ to $Y_{[N]}$. We prove that for all instances $Y_{[N]}$ it holds that
\begin{align*}
    \frac{\E\left(\ALG_t(Y_{[N]})\right)}{\E(\OPT(Y_{[N]}))}
    \geq \frac{\E\left(\ALG_t\left(Y_{[N]}^{+0}\right)\right)}{\E\left(\OPT\left(Y_{[N]}^{+0}\right)\right)},
\end{align*}
which immediately implies the result. Clearly, $\E(\OPT(Y_{[N]}))= \E(\OPT(Y_{[N]}^{+0}))$, as the arrival time of the added $0$ is independent of the other arrival times. We conclude by proving that $\E\left(\ALG_t\left(Y_{[N]}\right)\right)\geq \E\left(\ALG_t\left(Y_{[N]}^{+0}\right)\right)$. In fact, we can couple the arrival times of the values of $Y_{[N]}$ with the corresponding ones in $Y_{[N]}^{+0}$, and for the latter, add an independent arrival time for $0$. Since the $0$ is the smallest element, the relative rank of all other values is the same in both instances. Therefore, every time $\ALG_t$ selects a positive element in $Y_{[N]}^{+0}$, it selects the same element in $Y_{[N]}$. When $\ALG_t$ selects the $0$ in $Y_{[N]}^{+0}$, it may select a positive element in $Y_{[N]}$ or not stop at all. Thus, with this coupling we get that $\ALG_t(Y_{[N]}) \geq \ALG_t\left(Y_{[N]}^{+0}\right)$.
\end{proof}

\subsection{Proof of \cref{lem:convergent_prob_binom_model}}

\begin{proof}
For ease of notation, in what follows we write $\ALG_t$ instead of $\ALG_t(Y_{[N]})$. We have that
\begin{align*}
\PP(\ALG_t = Y_j) &= \int_p^1 \PP(\ALG_t = Y_j \,|\, Y_j \text{ arrives at time } \tau) \, d\tau\\
&= \sum_{i=1}^{\infty} \int_{t_i}^{t_{i+1}} \PP(\ALG_t = Y_j \,|\, Y_j \text{ arrives at time } \tau) \, d\tau \\
&= \sum_{i=1}^{\infty} \int_{t_i}^{t_{i+1}} \sum_{\ell=1}^{j\land i} \PP(\ALG_t \text{ does not stop before } \tau \, |\, Y_j \text{ is } \ell\text{-local and arrives at } \tau)\\
&{}\qquad \cdot \PP(Y_j \text{ is } \ell \text{-local}\,|\, Y_j \text{ arrives at } \tau ) \, d\tau\\
&= \sum_{i=1}^{\infty} \int_{t_i}^{t_{i+1}} \sum_{\ell=1}^{j\land i}
\PP(\ALG_t \text{ does not stop before } \tau \, |\, Y_j \text{ is } \ell\text{-local and arrives at } \tau)\\
&{}\qquad \cdot \binom{j-1}{\ell-1} (1-\tau)^{j-\ell} \tau^{\ell-1} \,d\tau .
\end{align*}
The last equality comes from the fact that $Y_j$ is $\ell$-local if exactly $\ell-1$ items from $Y_1,...,Y_{j-1}$ arrive before $Y_j$. Now, note that the event that $\ALG_t$ stops before $\tau$ does not depend on what elements arrive after $\tau$ and what are their relative rankings, but only on the relative rankings of the items that arrive before $\tau$. Also, note that when $N$ is large, the probability that at least $i$ items arrive before a given time $\tau>0$ tends to $1$. Therefore,
\begin{align*}
    &\PP(\ALG_t \text{ does not stop before } \tau\, |\, Y_j\text{ is } \ell\text{-local and arrives at } \tau)\\
    &= \PP( \ALG_t \text{ does not stop before } \tau\, |\, \text{at least } i \text{ items arrive before } \tau) + o(N)\\
    &= \prod_{r=1}^i \PP( r\text{-th largest item before } \tau \text{ arrives before } t_r) +o(N)\\
    &= \prod_{r=1}^i \frac{t_r}{\tau} + o(N) = \frac{T_i}{\tau^i} + o(N).
\end{align*}
Taking limit when $N$ tends to infinity we conclude the proof of the lemma.
\end{proof}

\subsection{Proof of Theorem~\ref{thm:convergence_to_limit}}
\label{sec:appendixA}

We first introduce two lemmas that bound the ratio between the coefficients of the linear programs. Then, to bound the difference between the values, we produce a solution for one problem from a solution to the other, and vice versa.

\begin{lem}
\label{lem:approximate_coeff_1}
For integers $N,h,k$, such that $p=\frac{h}{N}\in (0,1)$, $N\geq\frac{1}{(1-p)}+1$, and $1\leq k \leq h+1$, we have that
\begin{align}
    1\leq\frac{\sum\limits_{j=1}^k \frac{N-h}{N-j+1}\prod\limits_{s=0}^{j-2}\frac{h-s}{N-s}}{
    1-p^k} \leq 1+\frac{e}{(1-p)(N-1)} \,.
\end{align}
\end{lem}
\begin{proof}
Denote $A_j= \frac{N-h}{N-j+1}\prod\limits_{s=0}^{j-2}\frac{h-s}{N-s}$ and
$B_j= (1-p)p^{j-1}$. In order to bound the ratio $\frac{\sum_{j=1}^k A_j}{\sum_{j=1}^k B_j}$ for $1\leq k\leq h+1$, we find a uniform bound on $\frac{A_j}{B_j}$ for $1\leq j \leq h+1$. 

Recall that by definition $h= p\cdot N$. It is easy to see that $\frac{A_1}{B_1} =1$, and that $\frac{A_{j+1}}{B_{j+1}} = \frac{A_j}{B_j}\cdot \frac{h-j+1}{(N-j)\cdot p}$. Therefore, $\frac{A_{j+1}}{B_{j+1}}\geq \frac{A_j}{B_j}$ if and only if $h-j+1\geq h -pj$, which is equivalent to $j\leq\frac{1}{1-p}$. Then we can conclude that for all $j\leq h+1$,
\begin{align*}
    \frac{A_j}{B_j} 
    &= \frac{A_1}{B_1}\cdot \prod_{i=1}^{j-1} 
    \left( \frac{h-i+1}{h-pi} \right)\\
    &\leq \prod_{i=1}^{\lfloor 1/(1-p) \rfloor} 
    \left( \frac{h-i+1}{h-pi} \right)\\
    &\leq \left( \frac{h}{h-p} \right)^{\lfloor 1/(1-p) \rfloor} \\
    &= \left( 1 + \frac{1}{N-1} \right)^{\lfloor 1/(1-p) \rfloor}\\
    &\leq 1 + \frac{1}{N-1}\cdot \frac{1}{1-p} \left( 1+ \frac{1}{1/(1-p)} \right)^{1/(1-p)}\\
    &\leq 1+ \frac{e}{(1-p)(N-1)},
\end{align*}
where the second last inequality comes from doing a first-order approximation of a convex function.

For the lower bound of $1$, it is enough to note that $\sum_{j=1}^{h+1}A_j=1$ and that $\sum_{j=1}^{h+1} B_j\leq \sum_{j=1}^\infty B_j=1$, together with the already mentioned fact that $A_j\geq B_j$ if and only if $j\leq \frac{1}{1-p}$. \end{proof}

\begin{lem}
\label{lem:approximate_coeff_2}
For positive integers $N,i,j,\ell$ such that $N\geq 32$, $\frac{\sqrt{N}\log N}{1-p}\leq i \leq N-\frac{\sqrt{N}\log N}{1-p}$, $j\leq \frac{\log N}{1-p}$, and $\ell\leq j$, and for a real $t\in \left[ \frac{i-1}{N}, \frac{i}{N} \right]$, we have that
\begin{align}
    1-\frac{3\log N}{(1-p)\sqrt{N}} \leq \frac{\frac{i}{N}\frac{\binom{j-1}{\ell-1}\binom{N-j}{i-\ell}}{\binom{N-1}{i-1}}}{\binom{j-1}{\ell-1} (1-t)^{j-\ell}t^\ell}
    \leq 1+ \frac{5\log N}{(1-p) \sqrt{N}}\label{eq:coeff_ratio}
\end{align}
\end{lem}

\begin{proof}
We start by rewriting the expression in the middle of \cref{eq:coeff_ratio}.
\begin{align}
    \frac{\frac{i}{N}\frac{\binom{j-1}{\ell-1}\binom{N-j}{i-\ell}}{\binom{N-1}{i-1}}}{\binom{j-1}{\ell-1} (1-t)^{j-\ell}t^\ell}
    &= \frac{\frac{i}{N}\cdot \frac{(N-i)!}{(N-i-j+\ell)!} \cdot \frac{(i-1)!}{(i-\ell)!} \cdot \frac{(N-j)!}{(N-1)!}}{(1-t)^{j-\ell} t^\ell}\\
    &= \frac{\prod_{k=0}^{j-\ell-1} \frac{N-i-k}{N-k}\cdot \prod_{k=0}^{\ell-1} \frac{i-k}{N-j+\ell-k} }{(1-t)^{j-\ell} t^\ell}\,.
    \label{eq:lem_coeff_ratio_1}
\end{align}
Now, the expression in \cref{eq:lem_coeff_ratio_1} is clearly at most

\begin{align*}
    \frac{\prod_{k=0}^{j-\ell-1} \frac{N-i-k}{N-k}\cdot \prod_{k=0}^{\ell-1} \frac{i-k}{N-j+\ell-k} }{
    \left( \frac{N-i}{N} \right)^{j-\ell} \left( \frac{i-1}{N} \right)^{\ell}}
    &= \prod_{k=0}^{j-\ell-1}\left( \frac{N-i-k}{N-i}\cdot \frac{N}{N-k}\right)
    \cdot \prod_{k=0}^{\ell-1}\left( \frac{i-k}{i-1}\cdot \frac{N}{N-j+\ell-k}\right)\\
    &\leq \frac{i}{i-1}\cdot \left(\frac{N}{N-j}\right)^{j}\\
    &= \left(1+\frac{1}{i-1}\right)\cdot
    \left(1+\frac{j}{N-j}\right)^j\\
    &\leq \left(1+\frac{1}{\sqrt{N}\log_2N-1}\right)\cdot
    \left( 1+\frac{j^2 e}{N-j}\right)\\
    &\leq 1+\frac{5\log N}{(1-p)\sqrt{N}}
\end{align*}

And is also at least
\begin{align*}
    \frac{ (N-i-j+\ell)^{j-\ell} (i-\ell)^\ell \frac{1}{N^{j}} }{\left( \frac{N-i+1}{N} \right)^{j-\ell} \left( \frac{i}{N} \right)^{\ell}}
    &= \frac{(N-i-j+\ell)^{j-\ell}(i-\ell)^\ell}{(N-i+1)^{j-\ell} i^\ell}\\
    &= \left( 1- \frac{j-\ell+1}{N-i+1}\right)^{j-\ell} \left(1-\frac{\ell}{i} \right)^\ell\\
    &\geq 1-\frac{(j-\ell)(j-\ell+1)}{N-i+1}-\frac{\ell^2}{i}
    \\
    &\geq 1- 3\frac{\log_2 N}{(1-p)\sqrt{N}}\,.
\end{align*}
\end{proof}

\begin{proof}[Proof of \cref{thm:convergence_to_limit}]
To prove the theorem we take a solution to one problem and transform it into a solution of the other. Let $N,h$ be integers and $0<p<1$ a scalar such that $h=p\cdot N$. We start with an optimal solution $(q^*,\alpha(p))$ for $\SDCLP_p$, and define for a given $N\geq 1$ a solution $(x,\alpha')$ as follows.
\begin{align*}
    x_{i,\ell}&= \int_{\frac{i-1}{N}}^{\frac{i}{N}} q^*(t,\ell) dt,\; \text{ for } i\in[N]\setminus [h], \ell\in [i]\\
    \alpha'&= \min_{k\in [h+1]} 
    \frac{\displaystyle\sum\limits_{j=1}^k \displaystyle\sum\limits_{i=h+1}^{N} \displaystyle\sum\limits_{\ell=1}^{j}\frac{ix_{i,\ell}}{N}\frac{\binom{j-1}{\ell-1}\binom{N-j}{i-\ell}}{\binom{N-1}{i-1}}}{\sum\limits_{j=1}^k \frac{N-h}{N-j+1}\prod\limits_{s=0}^{j-2}\frac{h-s}{N-s}}
\end{align*}
We prove first that $(x,\alpha')$ is a feasible solution for $\SDLP_{pN,N})$. 
Note that for given $i\in [N]\setminus [h]$, $\ell\in [i]$, and $t\in [\frac{i-1}{N},\frac{i}{N}]$ we have from the feasibility of $q^*$ that
\begin{align*}
    tq^*(t,\ell) + \int_{\frac{i-1}{N}}^t q^*(\tau,\ell)d\tau
    &\leq 1- \int_{p}^{\frac{i-1}{N}} \sum_{s\geq 1} q^*(\tau,s)d\tau\\
    &\leq 1-\sum_{j=h+1}^{i-1}\sum_{s=1}^j x_{j,s}\,.
\end{align*}
Integrating on both sides we obtain that
\begin{align*}
    && \int_{\frac{i-1}{N}}^{\frac{i}{N}}\left(
    tq^*(t,\ell) + \int_{\frac{i-1}{N}}^t q^*(\tau,\ell)d\tau\right)\, dt
    &\leq 
    \int_{\frac{i-1}{N}}^{\frac{i}{N}}\left(
    1-\sum_{j=h+1}^{i-1}\sum_{s=1}^j x_{j,s}\right)\, dt\\
    \Leftrightarrow &&
    \left. t\int_{\frac{i-1}{N}}^tq^*(\tau,\ell)d\tau \right|_{t=\frac{i-1}{N}}^{t=\frac{i}{N}}
    &\leq \frac{1}{N}\left(
    1-\sum_{j=h+1}^{i-1}\sum_{s=1}^j x_{j,s}\right)\\
    \Leftrightarrow &&
    i\cdot x_{i,\ell} 
    &\leq 1-\sum_{j=h+1}^{i-1}\sum_{s=1}^j x_{j,s}\,,
\end{align*}
where in the second inequality we applied integration by parts on the left-hand side. Therefore, $x$ is a feasible solution. We now give an upper bound for $\alpha(p)-\alpha'$. From the definition of $\alpha'$ and \cref{lem:approximate_coeff_1}, together with the fact that $1/(1+y)\geq 1-y$ for all $y\geq 0$, we obtain that
\begin{align*}
    \alpha'\geq \min_{k\in [h+1]} 
    \frac{\displaystyle\sum\limits_{j=1}^k \displaystyle\sum\limits_{i=h+1}^{N} \displaystyle\sum\limits_{\ell=1}^{j}\frac{ix_{i,\ell}}{N}\frac{\binom{j-1}{\ell-1}\binom{N-j}{i-\ell}}{\binom{N-1}{i-1}}}{1-p^k}\cdot\left(
    1-\frac{e}{(1-p)(N-1)}\right).
\end{align*}
Now, if $k>\frac{\log N}{1-p}\geq \log_{p}(1/N)$, then $p^k\leq 1/N$, so we can take in the minimization $k\leq \frac{\log N}{1-p}$ and lose a factor $(1-1/N)$. Denote $i^*=\frac{\sqrt{N}\log N}{1-p}$.
Since $j\leq k$, after replacing $x_{i,\ell}$ with the integral that defines it, we can apply \cref{lem:approximate_coeff_2} to obtain that
\begin{align*}
    \alpha'&\geq \min_{1\leq k\leq \frac{\log N}{1-p}}
    \frac{\displaystyle\sum_{j=1}^k
    \sum_{i=(h+1)\lor i^*}^{
    N-i^*}
    \sum_{\ell=1}^j \int_{\frac{i-1}{N}}^{\frac{i}{N}}
    q^*(t,\ell)\binom{j-1}{\ell-1} (1-t)^{j-\ell} t^\ell dt
    }{1-p^k}\cdot\left(1-\frac{7\log N}{(1-p)\sqrt{N}}
    \right)\\
    &=\min_{1\leq k\leq \frac{\log N}{1-p}}
    \frac{\displaystyle\sum_{j=1}^k
    \int_{p\lor\frac{i^*}{N}}^{1-\frac{i^*}{N}}
    \sum_{\ell=1}^j
    q^*(t,\ell)\binom{j-1}{\ell-1} (1-t)^{j-\ell} t^\ell dt
    }{1-p^k}\cdot\left(1-\frac{7\log N}{(1-p)\sqrt{N}}
    \right)\,.
\end{align*}
Now, since $t\cdot q^*(t,\ell)\leq 1$ for all $\ell,t$ and $\sum_{\ell=1}^j \binom{j-1}{\ell-1}(1-t)^{j-\ell}t^{\ell-1}=1$ for all $j\geq 1$, we get that
\begin{align*}
    \alpha'&\geq \min_{1\leq k\leq \frac{\log N}{1-p}}
    \frac{\displaystyle\sum_{j=1}^k
    \int_{p}^{1}
    \sum_{\ell=1}^j
    q^*(t,\ell)\binom{j-1}{\ell-1} (1-t)^{j-\ell} t^\ell dt
    }{1-p^k}\cdot\left(1-\frac{7\log N}{(1-p)\sqrt{N}}
    - \frac{2i^*k}{N}\right)\\
    &\geq \alpha(p)\cdot\left(
    1-\frac{9(\log N)^2}{(1-p)^2\sqrt{N}}\right)\\
    &\geq \alpha(p)-\frac{9(\log N)^2}{(1-p)^2\sqrt{N}}\,.
\end{align*}
We prove now the other side of the inequality. Let $(x^*,\alpha_{N,p})$ be an optimal solution for $\SDLP_{pN,N}$. We construct a solution $(q,\alpha'')$ as follows.
\begin{align*}
    q(t,\ell)&= \begin{cases}
    N x^*_{i,\ell}\cdot \left(
    1-\frac{\log N}{(1-p)\sqrt{N}}\right) \,, \;\;&\text{ for } t\in [p,1], \text{ if } i= \lceil t\cdot N \rceil\geq \sqrt{N} \text{ and } \ell\leq i\land \frac{\log N}{1-p}\\
    0\,\;\; &\text{ for } t\in [p,1],
    \text{ if } i=\lceil t\cdot N \rceil< \sqrt{N} \text{ or }  \ell> i\land \frac{\log N}{1-p}
    \end{cases}\\
    \alpha''&= \min_{k\geq 1}
    \frac{\sum_{j=1}^k
    \int_p^1 \sum_{\ell=1}^j
    q(t,\ell)\binom{j-1}{\ell-1}(1-t)^{j-\ell}
    t^\ell dt}{1-p^k}\,.
\end{align*}
Now, we can check this solution is feasible in the continuous problem. In fact, for $t< \frac{1}{\sqrt{N}}$, it is trivially satisfied because $q(t,\ell)=0$ for all $\ell$. For $t\geq p\lor \frac{1}{\sqrt{N}}$, $i=\lceil t\cdot N\rceil$, and any $\ell\geq 1$,
\begin{align*}
    tq(t,\ell) + \int_p^t \sum_{s\geq 1} q(\tau,s)d\tau 
    &\leq \left(i x^*_{i,\ell} + \sum_{j=h+1}^{i-1}\sum_{s=1}^j x^*_{j,s} +
    \int_{\frac{i-1}{N}}^t \sum_{s=1}^{\frac{\log N}{1-p}} Nx^*_{i,s} d\tau\right)\cdot
    \left( 1-\frac{\log N}{(1-p)\sqrt{N}}\right)\\
    &\leq\left( 1 + \frac{\log N}{i(1-p)}\right)\cdot
    \left( 1-\frac{\log N}{(1-p)\sqrt{N}}\right)\\
    &\leq \left(1+ \frac{\log N}{(1-p)\sqrt{N}}\right)\cdot
    \left( 1-\frac{\log N}{(1-p)\sqrt{N}}\right)\\
    &\leq 1\,,
\end{align*}
where the first inequality comes from replacing with the definition of $q$, and the third one comes from the fact that $i\geq \sqrt{N}$.

We argue similarly to the lower bound for $\alpha'$, using \cref{lem:approximate_coeff_1,lem:approximate_coeff_2}, together with the extra factor $
    \left( 1-\frac{\log N}{(1-p)\sqrt{N}}\right)$ that was necessary for the feasibility constraint. This yields the inequality
\begin{align*}
    \alpha''&\geq \alpha_{N,p}
    - \frac{5(\log N)^2}{(1-p)^2\sqrt{N}}
    - \frac{\log N}{(1-p) \sqrt{N}}\\
    &\geq \alpha_{N,p} - \frac{6(\log N)^2}{(1-p)^2 \sqrt{N}}
    \,.
\end{align*}

\end{proof}

\end{document}